\documentclass[a4paper,11pt]{article}   %%FRK: eqsecnum does not work, see below
\usepackage{pos}
\usepackage{amsfonts}
\usepackage{dsfont}  %%FRK needed for \id below

\newcommand{\version}{v5} %%true version v4991  %%FRK
\newcommand{\beq}{\begin{equation}}
\newcommand{\eeq}{\end{equation}}
\newcommand{\beqa}{\begin{eqnarray}}
\newcommand{\eeqa}{\end{eqnarray}}
\newcommand{\bsubeqs}{\begin{subequations}}
\newcommand{\esubeqs}{\end{subequations}}

\def\id{\mathds{1}}   %%FRK use 'double-stroke' symbol

\title{IIB matrix model, bosonic master field, and emergent spacetime}
\author*[a]{F.R. Klinkhamer}
\affiliation[a]{Institute for Theoretical Physics,
Karlsruhe Institute of Technology (KIT),\\
76128 Karlsruhe, Germany}

\emailAdd{frans.klinkhamer@kit.edu}
\abstract{The IIB matrix model has been suggested as a particular formulation
of nonperturbative superstring theory (M-theory).
It has now been realized that an emerging classical spacetime may
reside in its large-$N$ master field.
This bosonic master field can, in principle, give rise to Minkowski
and Robertson-Walker spacetimes.
The outstanding task is to solve the bosonic master-field equation,
which is essentially an algebraic equation.
In this article, we present new results for
the $(D,\,N)=(10,\,4)$ bosonic master-field equation of the
IIB matrix model,
where $D$ is the number of bosonic matrices and $N$ the matrix size.
We also give, in a self-contained appendix, explicit results for
critical points of the effective bosonic action.
The main physics application of the (dimensionless) IIB matrix model
may be in providing a (conformal) phase that replaces the
Friedmann big bang singularity.}

\FullConference{%
  Corfu Summer Institute 2021 "School and Workshops on Elementary Particle Physics and Gravity"\\
  29 August - 9 October 2021\\
  Corfu, Greece
\hfill preprint\;arXiv:2203.15779 \;(\version)
}

\tableofcontents

\numberwithin{equation}{section}
\begin{document}

\maketitle

\section{Introduction}
\label{sec:Introduction}

One of the great questions of physics is:
how did the Universe start?
%what happened at the birth of the Universe?
Or, at a more technical level:
what replaces the big bang singularity of
the Friedmann cosmology~\cite{Friedmann1922-1924}?
The Friedmann big bang singularity would have infinite energy density and
curvature, but it is clear that Nature somehow will prevent
these infinities from occurring.

Recall that general relativity~\cite{Einstein1916}
and the standard model of elementary particle physics~\cite{Lee1981}
are the current underlying theories of
the standard Friedmann cosmology~\cite{Friedmann1922-1924}.
Hence, the answer to the second question above, most likely, requires an extended theory,
which incorporates general relativity and the standard model.
In this article,
we will consider superstrings~\cite{GreenSchwarzWitten1987} as such a candidate theory.
We will need, in fact, a nonperturbative formulation
of superstring theory.

The final formulation of nonperturbative superstring
theory, also known as $M$-theory~\cite{Witten1995,HoravaWitten1996},
is still incomplete.
The IKKT matrix model~\cite{IKKT-1997} is one suggestion.
As that matrix model reproduces the basic structure
of the light-cone string field theory of type-IIB superstrings,
the model is also known as the IIB matrix
model~\cite{Aoki-etal-review-1999}. It may, therefore,
be a worthwhile undertaking to investigate the IIB matrix model thoroughly.

A few years ago, we started thinking about how the
IIB matrix model could, in principle, produce
a new phase replacing the Friedmann big bang singularity,
with an emerging classical spacetime and emergent matter.
But how the spacetime, in particular, emerged from the
matrix model was far from clear. We then realized
that the large-$N$ master field of
Witten~\cite{Witten1979,Coleman1985,GreensiteHalpern1983,Carlson-etal1983}
could play a crucial role and we showed that, in principle,
the master field of the IIB matrix model could give rise
to the points and the metric of an emerging classical
spacetime~\cite{Klinkhamer2020-master}. This discussion
was, however, based on  \emph{assumed} master-field matrices
and we really need to \emph{calculate} them. Our work
of the last year has focussed on that task.

For the first conceptual phase of our endeavours,
with a main paper~\cite{Klinkhamer2020-master}
and a cosmology follow-up paper~\cite{Klinkhamer2020-reg-bb-IIB-m-m}
available,
we have already written a review~\cite{Klinkhamer2021-APPB-review}.
For the second calculational phase, with three technical
papers~\cite{Klinkhamer2021-first-look,%
Klinkhamer2021-sols-D3-N3,Klinkhamer2021-numsol} available,
we now wish to present a corresponding review that focuses
on the main results, while leaving out the technicalities.
We also take the opportunity to present some new numerical results.

The outline of our paper is as follows.
In Sec.~\ref{sec:IIB-matrix-model}, we review
the main points of the IIB matrix model.
In Sec.~\ref{sec:Bosonic-master-field-and-master-field-equation},
we introduce the bosonic master field
and the equation which determines it.
In Sec.~\ref{sec:Numerical-solutions-D10-N4},
we present numerical solutions for the $(D,\,N)=(10,\,4)$ case,
both with and without dynamical fermions
(here, $D$ is the number of bosonic matrices and $N$ the matrix size).
In Sec.~\ref{sec:Conclusion}, we give some closing remarks,
also about getting a ``tamed'' big bang
from the IIB matrix model.
There are furthermore five appendices with technical details and
additional results.
In App.~\ref{app:Large-N-factorization}, we
present a few new results in favor of the crucial property
of large-$N$ factorization in the IIB matrix model.
In App.~\ref{app:SU4-generators}, we give the explicit
choice for the $SU(4)$ generators used in our calculations.
In App.~\ref{app:Pseudorandom-numbers}, we list one particular
realization of pseudorandom numbers entering the
$(D,\,N)=(10,\,4)$ master-field equation, which is essentially an
algebraic equation.
In App.~\ref{app:Coefficients-from-full-algebraic-equation},
we give the obtained coefficients of an
approximate numerical solution of the full $(D,\,N)=(10,\,4)$
algebraic equation.
In App.~\ref{app:Nontrivial-critical-points}, we
look for solutions of another algebraic equation, the
stationarity equation from the effective bosonic action, and present
two nontrivial critical points for the case $(D,\,N)=(3,\,3)$.

%%\newpage%%tmp
\section{IIB matrix model}
\label{sec:IIB-matrix-model}

%%based on sols-bos-master-eq-v5.tex final version APPB-52-1339(2021)

\subsection{Partition function}
\label{subsec:Partition-function}

Let us recall the definition of the IIB matrix
model~\cite{IKKT-1997,Aoki-etal-review-1999}.
We essentially take over the conventions and notations of
Ref.~\cite{KrauthNicolaiStaudacher1998},
except that we write $A^{\mu}$ for the bosonic matrices
with a directional index $\mu$ running over $\{1,\,\ldots\,,D\}$
for $D=10$.
These bosonic matrices $A^{\mu}$, as well as the
fermionic matrices $\Psi_{\alpha}$\,,
are $N \times N$ traceless Hermitian matrices.
The partition function of the supersymmetric IIB matrix model (IIB-m-m)
for $N\geq 2$ is then defined as
follows~\cite{IKKT-1997,Aoki-etal-review-1999,KrauthNicolaiStaudacher1998}:
\bsubeqs\label{eq:Z-D-N-F-all-defs}
\beqa
\label{eq:Z-D-N-F}
\hspace*{-8.0mm}
Z_{D,\,N}^{F} &=&
\int \prod_{c=1}^{g}\,\prod_{\mu=1}^{D}\,
d A_{\mu}^{c}\;
%%\frac{d A_{\mu}^{c}}{\sqrt{2\pi}}\;
%%\exp\left( -\,S_{\text{bos}}[A] \right)\;
e^{\displaystyle{-\,S_{\text{bos}}}[A]}\,
%%\nonumber\\\hspace*{-8.0mm}&& \times
\left(\int \prod_{c=1}^{g}\,\prod_{\alpha=1}^{\mathcal{N}}
d\Psi_{\alpha}^{c}\;
%%\exp\left( -\,S_{\text{ferm}}[A,\,\Psi] \right)
e^{\displaystyle{-\,S_{\text{ferm}}[A,\,\Psi]}}\,
\right)^{F}\!,
\\[1.05mm]
\hspace*{-8.0mm}
\label{eq:Sbos-IIB}
S_{\text{bos}}[A] &=&
-\frac{1}{2}\,\text{Tr}\,
\Big(\big[A^{\mu},\,A^{\nu} \big]\,\big[ A^{\mu},\,A^{\nu} \big]\,\Big)\,,
%%\eeqa
\\[1.05mm]
%%\beqa
\hspace*{-8.0mm}
\label{eq:Sferm-IIB}
S_{\text{ferm}}[A,\,\Psi] &=&
-\text{Tr}\,
\Big(\, \Psi_{\alpha}\,\left(\mathcal{C}\,\Gamma^{\mu}\right)_{\alpha\beta}\,
\big[ A^{\mu},\,\Psi_{\beta}\big]\Big)\,,
%%\eeqa
\\[1.05mm]
%%\beqa
\hspace*{-8.0mm}
\label{eq:Amu-coeff-Psialpha-coeff}
A_{\mu}
&=&A_{\mu}^{c}\,t_{c}\,,
\quad
\Psi_{\alpha} = \Psi_{\alpha}^{c}\,t_{c}\,,
\quad
t_{c} \in \text{su}(N)\,,
\\[1.05mm]
\hspace*{-8.0mm}
\label{eq:trace-t-t}
\text{Tr}\, \big( t_{c} \cdot t_{d} \big)
&=& \frac{1}{2}\;\delta_{c d}\,,
%%\eeqa
\\[1.05mm]
%%\beqa
\label{eq:g}
\hspace*{-8.0mm}
g &\equiv& N^2-1\,,
\\[1.05mm]
\label{eq:mathcalN}
\hspace*{-8.0mm}
\mathcal{N} &=& 2\,\big(D-2\big)=
%%2,\,4,\,8,\,16 \,,
2,\,4,\,16 \,,
\quad
%%\text{for}\;\;D=3,\,4,\,6,\,10 \,,
\text{for}\;\;D=3,\,4,\,10 \,,
\\[1.05mm]
\label{eq:F-0-or-1}
\hspace*{-8.0mm}
F &\in&  \big\{ 0,\, 1 \big\}\,,
\\[1.05mm]
\label{eq:D10-F1}
\hspace*{-8.0mm}
\big\{D,\, F \big\}\Big|^{\text{(IIB-m-m)}}
 &=& \big\{10,\, 1 \big\}\,,
\eeqa
\esubeqs
where repeated Greek indices are summed over
(just as with an implicit Euclidean ``metric'') and
$F$ is an on/off parameter to include ($F=1$)
or exclude ($F=0$) dynamic-fermion effects.
The commutators entering %%the action terms
\eqref{eq:Sbos-IIB}
and \eqref{eq:Sferm-IIB} are defined
by $[X,\,Y]\equiv$ $X \cdot Y - Y \cdot X$
for square ma\-tri\-ces $X$ and $Y$ of equal dimension.
The fermionic action \eqref{eq:Sferm-IIB}
contains also the charge conjugation matrix $\mathcal{C}$
and the $\Gamma^{\mu}$ for $D=10$ are
Weyl-projected ``gamma'' matrices, $\Gamma^{\mu}=\Sigma^{\mu}$.
These $16\times 16$ matrices $\Sigma^{\mu}$ have been given explicitly in
Refs.~\cite{NishimuraVernizzi2000-JHEP,Klinkhamer2021-numsol}
and their notation recalls the $2\times 2$  Pauli matrices  $\sigma^{\mu}$
of the four-dimensional case with $4\times 4$  Dirac matrices  $\gamma^{\mu}$
in the chiral representation.

The expansions \eqref{eq:Amu-coeff-Psialpha-coeff},
with real coefficients $A_{\mu}^{c}$ and
real Grassmannian coefficients $\Psi_{\alpha}^{c}$,
use the $N \times N$ traceless Hermitian
$SU(N)$ generators $t_{c}$ with normalization \eqref{eq:trace-t-t}
and structure constants $f^{abc}$ as given by \eqref{eq:structure-constants}
in App.~\ref{subapp:Supersymmetric-model-D4-N4}.
(The matrix model for $D=6$ has Weyl spinors without Majorana condition,
while the model has Majorana spinors for $D=3,4$
and Majorana--Weyl spinors for $D=10$;
see App.~4.~A of Ref.~\cite{GreenSchwarzWitten1987} for a brief review of
supersymmetric Yang--Mills gauge theory in $D=3,\,4,\,6,\,10$ spacetime dimensions.)
Equation \eqref{eq:mathcalN} corrects the corresponding equation
in Ref.~\cite{Klinkhamer2021-sols-D3-N3}.

The action of the model \eqref{eq:Z-D-N-F-all-defs}
is invariant under the following global gauge transformation:%
\begin{eqnarray}
\label{eq:IIB-matrix-model-global-gauge-transformation}
A^{\,\mu} &\to&  \Omega\, A^{\,\mu}\,\Omega^{\dagger}\,,
\quad
\Psi_{\alpha} \to \Omega\, \Psi_{\alpha}\,\Omega^{\dagger}\,,
\quad
\Omega \in  SU(N)\,.
\end{eqnarray}
In addition, there is an $SO(D)$ rotation invariance and
supersymmetry~\cite{IKKT-1997,Aoki-etal-review-1999}.
Note that, as it stands,
the model variables $A_{\mu}^{c}$ and $\Psi_{\alpha}^{c}$
in \eqref{eq:Z-D-N-F-all-defs}
are dimensionless (see Sec.~\ref{subsec:Conceptual-remarks}
for further remarks).%

%%%%\newpage%%tmp
The Gaussian-type integrals over the Grassmann variables $\Psi_{\alpha}^{c}$
in \eqref{eq:Z-D-N-F} can be performed analytically,
so that the partition function reduces to a purely bosonic integral,
\bsubeqs\label{eq:Z-D-N-F-with-Pfaffian-and-Seff-D-N-F}
\beqa
\label{eq:Z-D-N-F-with-Pfaffian}
\hspace*{-12.0mm}
&&
Z_{D,\,N}^{F} =%%&=&
\int \prod_{c=1}^{g}\,\prod_{\mu=1}^{D}\,
d A_{\mu}^{c}\;
e^{\displaystyle{-\,S_{\text{bos}}[A]}}\;
\Big( \mathcal{P}_{D,\,N}[A] \Big)^{F}
%%\nonumber\\[1mm]&=&
=
\int \prod_{c=1}^{g}\,\prod_{\mu=1}^{D}\,
d A_{\mu}^{c}\;
e^{\displaystyle{-\,S_{\text{eff},\,D,\,N}^{F}[A]}}\,,
\\[1mm]
\label{eq:Seff-D-N-F}
\hspace*{-12.0mm}
&&
S_{\text{eff},\,D,\,N}^{F}[A]
=S_{\text{bos}}[A]- F\;\log\,\mathcal{P}_{D,\,N}[A]\,.
\eeqa
\esubeqs
The obtained quantity $\mathcal{P}_{D,\,N}[A]$
corresponds, in fact, to the following
Pfaffian~\cite{KrauthNicolaiStaudacher1998,NishimuraVernizzi2000-JHEP}:
\bsubeqs\label{eq:calP-calM}
\beqa
\label{eq:calP}
&&
\mathcal{P}_{D,\,N}[A] = \text{Pf}\left[\mathcal{M}\left(A\right)\right]\,,
\\[2mm]
\label{eq:calM}
&&
\mathcal{M}_{a\alpha\,,\,b\beta}
 =
-i\, f_{abc}\, \big(\mathcal{C}\,\Gamma_{\mu}\big)_{\alpha\beta} \;A_{\mu}^{\;c}
\equiv
\mathcal{M}_{\underline{A}\,,\,\underline{B}}\,,
\eeqa
\esubeqs
with Lie-algebra indices $a,b,c$ running over $\{1,\, \ldots \,,\, g\}$,
spinorial indices $\alpha,\beta$ running
over $\{1,\, \ldots \,,\,\mathcal{N} \}$,
and the directional index $\mu$ being summed
over $\{  1,\,  \ldots \,,\,D \}$,
where the pair of indices $a\alpha$ gives a combined index $\underline{A}$
and the pair $b\beta$ a combined index $\underline{B}$.
Note that the matrix $\mathcal{M}_{\underline{A}\,,\,\underline{B}}$ is antisymmetric,
because $f_{abc}$ is antisymmetric in the indices $a,b$ and
$\big(\mathcal{C}\,\Gamma_{\mu}\big)_{\alpha\beta}$ symmetric in the indices $\alpha,\beta$.
Recall that the Pfaffian can be defined as
a sum over permutations~\cite{KrauthNicolaiStaudacher1998}
or as a sum involving the completely antisymmetric Levi--Civita symbol $\epsilon$
normalized to unity~\cite{Klinkhamer2021-numsol}.
This last definition of the Pfaffian
of a $(2 n)\times (2 n)$ skew-symmetric matrix $S=(s_{ij})$ reads
\beq
\label{eq:def-Pfaffian}
\text{Pf}[S]\equiv
\frac{1}{2^{n}\,n!}\;
\epsilon_{i_{1}j_{1} i_{2}j_{2} \cdots i_{n}j_{n} }\;
s_{i_{1}j_{1}}\,s_{i_{2}j_{2}} \cdots s_{i_{n}j_{n}}\,,
\eeq
with implicit summations over repeated indices.
An example is given by
the $2\times 2$ skew-symmetric  matrix $T=a\,i\,\sigma^{2}=
\big\{ \{0,\, a\} ,\,\{-a,\, 0\} \big\}$, which has the Pfaffian $\text{Pf}[T]=a$.

From \eqref{eq:calP-calM} and \eqref{eq:def-Pfaffian}, it is clear that
the Pfaffian $\mathcal{P}_{D,\,N}[A]$
is a homogenous polynomial in the bosonic coefficients
$A_{\mu}^{c}$, where the order $K$ is given by the following expression:
\beq
\label{eq:Pfaffian-order-polynomial-K}
K = \frac{1}{2}\,\mathcal{N}\, g=\big( D-2\big)\, \big( N^2-1\big)\,,
\quad
%%\text{for}\;\; D=3,\,4,\,6,\,10\,,
\text{for}\;\; D=3,\,4,\,10\,.
\eeq
%%including the $D=6$ case with complex Weyl spinors, as mentioned above.
Further discussion of the Pfaffian appears in, e.g.,
Refs.~\cite{KrauthNicolaiStaudacher1998,NishimuraVernizzi2000-JHEP,%
AustingWheater2001} and the $D=10$ Pfaffian has been detailed
in App.~A of Ref.~\cite{Klinkhamer2021-numsol}.

As mentioned above, the partition function of the genuine
IIB matrix model~\cite{IKKT-1997,Aoki-etal-review-1999}
has the following parameters in \eqref{eq:Z-D-N-F-all-defs}:
\beq
\label{eq:D-N-F-for-IIB-matrix-model}
D=10 \,, \quad F  =  1 \,, \quad  N \gg 1 \,,
\eeq
and there are now two supersymmetry transformations
(see Sec.~\ref{subsec:Conceptual-remarks} below).
The large-$N$ limit may require further discussion, but, at this
moment, we just consider $N$ to be large and finite
(for exploratory numerical results, see, e.g.,
Refs.~\cite{KimNishimuraTsuchiya2012,NishimuraTsuchiya2019,%
Anagnostopoulos-etal-2020} and references therein).

%%\newpage%%tmp
\subsection{Bosonic observables and expectation values}
\label{subsec:Bosonic-observables}

As our main interest is in the possible recovery of an emerging
classical spacetime, we primarily consider the bosonic observable
\beqa \label{eq:IIB-matrix-model-w-observable}
\hspace*{-0.00mm}
w^{\mu_{1} \,\ldots\, \mu_{m}}
&\equiv&
\frac{1}{N}\;
\text{Tr}\,\big( A^{\mu_{1}} \cdots\, A^{\mu_{m}}\big)\,,
\eeqa
which is invariant under \eqref{eq:IIB-matrix-model-global-gauge-transformation} and
where the $1/N$ prefactor on the right-hand side is solely for convenience.
Now, arbitrary strings of these bosonic observables
have expectation values
\beqa \label{eq:IIB-matrix-model-w-product-vev}
\hspace*{-2.0mm}
&&
\langle
w^{\mu_{1}\,\ldots\,\mu_{m}}\:w^{\nu_{1}\,\ldots\,\nu_{n}}\, \cdots\,
w^{\omega_{1}\,\ldots\,\omega_{z}}
\rangle_{D,\,N}^{F}
\nonumber\\[1mm]
\hspace*{-2.0mm}
&&
= \frac{1}{Z_{D,\,N}^{F}}\,\int dA\,
\big(w^{\mu_{1}\,\ldots\,\mu_{m}}\:w^{\nu_{1}\,\ldots\,\nu_{n}}\, \cdots\,
w^{\omega_{1}\,\ldots\,\omega_{z}}\big)\,
%%\exp\left(-\,S_{\text{eff},\,D,\,N}^{F}  \,\right)
e^{\displaystyle{-\,S_{\text{eff},\,D,\,N}^{F}}}\,,
\eeqa
where ``$dA$'' is a short-hand notation of the measure appearing
in \eqref{eq:Z-D-N-F} and $Z_{D,\,N}^{F}$ is defined by the
integral \eqref{eq:Z-D-N-F-with-Pfaffian}.

In closing, we emphasize that the IIB matrix model is relatively
straightforward to formulate,
having only a finite number of matrices, but hard to evaluate and interpret.

%%\newpage%%tmp
\section{Bosonic master field and master field equation}
\label{sec:Bosonic-master-field-and-master-field-equation}

%%based on sols-bos-master-eq-v5.tex final version APPB-52-1339(2021)

\subsection{Bosonic master field}
\label{subsec:Bosonic-master-field}

The expectation values \eqref{eq:IIB-matrix-model-w-product-vev},
at large values of $N$,
have a remarkable factorization property:
\beqa \label{eq:IIB-matrix-model-w-product-vev-factorized}
\hspace*{-0mm}
\langle
w^{\mu_{1}\,\ldots\,\mu_{m}}\:w^{\nu_{1}\,\ldots\,\nu_{n}}\, \cdots\,
w^{\omega_{1}\,\ldots\,\omega_{z}} \rangle_{D,\,N}^{F}
&\stackrel{N}{=}&
\langle w^{\,\mu_{1}\,\ldots\,\mu_{m}}\rangle_{D,\,N}^{F}\;
\langle w^{\,\nu_{1}\,\ldots\,\nu_{n}}\rangle_{D,\,N}^{F}
%%\nonumber\\[1mm]\hspace*{-00mm}&&
\, \cdots\,
\langle w^{\,\omega_{1}\,\ldots\,\omega_{z}}\rangle_{D,\,N}^{F}\,,
\nonumber\\[1mm]\hspace*{-00mm}&&
\eeqa
where the equality holds to leading order in $N$.
Evidence for the factorization property
\eqref{eq:IIB-matrix-model-w-product-vev-factorized} in the
context of the IIB matrix model has been presented in
Ref.~\cite{Ambjorn-etal2000}.
See also App.~\ref{app:Large-N-factorization} here for
further results in support of large-$N$ factorization.

According to Witten~\cite{Witten1979}, the factorization
(\ref{eq:IIB-matrix-model-w-product-vev-factorized}) implies that
the path integrals (\ref{eq:IIB-matrix-model-w-product-vev}) are
saturated by a single configuration,
which has been called the ``master field''~\cite{Coleman1985}
and whose matrices will be denoted by $\widehat{A}^{\,\mu}$.
To leading order in $N$, the expectation values
\eqref{eq:IIB-matrix-model-w-product-vev} are then given by
the bosonic master-field matrices $\widehat{A}^{\,\mu}$
in the following way:%
\bsubeqs \label{eq:IIB-matrix-model-w-product-vev-from-master-field}
\beqa
\hspace*{-4.00mm}
&&\langle
w^{\mu_{1}\,\ldots\,\mu_{m}}\:w^{\nu_{1}\,\ldots\,\nu_{n}}\, \cdots\,
w^{\omega_{1}\,\ldots\,\omega_{z}} \rangle_{D,\,N}^{F}
\;\stackrel{N}{=}\;
\widehat{w}^{\,\mu_{1}\,\ldots\,\mu_{m}}\:
\widehat{w}^{\,\nu_{1}\,\ldots\,\nu_{n}}\, \cdots\,
\widehat{w}^{\,\omega_{1}\,\ldots\,\omega_{z}},
\\[2mm]
\label{eq:def-what}
\hspace*{-4.00mm}&&
\widehat{w}^{\,\mu_{1}\,\ldots\, \mu_{m}}
\equiv
\frac{1}{N}\;
\text{Tr}\,\big( \widehat{A}^{\,\mu_{1}} \cdots\, \widehat{A}^{\,\mu_{m}}\big),
\eeqa
\esubeqs
where the master-field matrices $\widehat{A}^{\,\mu}$
have an implicit dependence on the model parameters
$D$, $N$, and $F$. Note that, for simplicity,
we speak about a single master field but there can be many~\cite{Carlson-etal1983}.
%
%In principle, it is possible that there is more than one master field,
%as long as these master fields give, in the large-N limit,
%exactly the same results for all possible observables of the type (5). %For simplicity, we will talk, in the following, about a single master field.

The task at hand is to obtain an equation which governs
these master-field matrices $\widehat{A}^{\,\mu}$
and to find solutions of that equation.

%%\newpage%%tmp
\subsection{Bosonic master-field equation}
\label{subsec:Bosonic-master-field-equation}

Introducing $N$ random constants $\widehat{p}_{k}$
and the $N \times N$ diagonal matrix
\beq
\label{eq:def-D}
D_{(\widehat{p})}(\tau)  \equiv
\text{diag}
\left(
%\exp\big[i\,\widehat{p}_{1}\,\tau\big],\, \,\ldots\, ,\,
%\exp\big[i\,\widehat{p}_{N}\,\tau\big]
e^{\displaystyle{i\,\widehat{p}_{1}\,\tau}},\, \,\ldots\, ,\,
e^{\displaystyle{i\,\widehat{p}_{N}\,\tau}}
\right)\,,
\eeq
the bosonic master-field matrices take
the following ``quenched'' form~\cite{GrossKitazawa1982}:
\bsubeqs\label{eq:Ahatrho-with-D-algebraic-equation-with-D}
\beqa\label{eq:Ahatrho-with-D}
\widehat{A}^{\;\mu}
&=&
D_{(\widehat{p})}(\tau_\text{eq})
\cdot \widehat{a}^{\;\mu}
\cdot D_{(\widehat{p})}^{-1}(\tau_\text{eq})\,,
\eeqa
for a sufficiently large value of $\tau_\text{eq}$
(see below for further explanations).
The $\tau$-independent matrix $\widehat{a}^{\;\mu}$
in \eqref{eq:Ahatrho-with-D} is determined by the
equation~\cite{GreensiteHalpern1983,Klinkhamer2020-master}
\beqa
\label{eq:algebraic-equation-with-D}
\frac{d}{d \tau}\,
\Big[D_{(\widehat{p})}(\tau)
\cdot \widehat{a}^{\;\mu}
\cdot D_{(\widehat{p})}^{-1}(\tau)\Big]_{\tau=0}
&=&
%-\left.\frac{\delta S_{\text{eff},\,D,\,N}^{F}}{\delta A_{\mu}}
%\right|_{A=\widehat{a}}\;
-\,\frac{\delta S_{\text{eff},\,D,\,N}^{F}\big[\,\widehat{a}\,\big]}
      {\delta\, \widehat{a}_{\mu}}\;
+\widehat{\eta}^{\;\mu}\,.
\eeqa
\esubeqs
All matrix indices have been suppressed in the three equations above
and $S_{\text{eff},\,D,\,N}^{F}[\,\widehat{a}\,]$
is given by \eqref{eq:Seff-D-N-F}.

%%%%%%%%%%%%%%%%%%%%%%%%%
%num-sol-master-eq-v5.tex final version  APPB-53-1-A5(2022)

It is instructive to write out \eqref{eq:algebraic-equation-with-D}
explicitly and to add matrix indices $\{k,\,l\}$
running over $\{  1,\,  \ldots \,,\,N \}$.
It is then clear that the equation is algebraic.
In fact, this algebraic equation
for $D$ traceless Hermitian matrices $\widehat{a}^{\;\mu}$
of dimension $N \times N$ reads
%\begin{subequations}
%\label{eq:full-algebraic-equation}
\begin{eqnarray}
\label{eq:full-algebraic-equation}
\hspace*{-6mm}&&
i\,\big(\widehat{p}_{k}-\widehat{p}_{l}\big)\;
\widehat{a}^{\;\mu}_{\;kl}
%&=&
+ \delta_{\lambda\nu}\,
\Big[\widehat{a}^{\,\lambda},\,\big[\widehat{a}^{\,\nu},\,
\widehat{a}^{\;\mu}\big]\Big]_{kl}
- F\,\frac{1}{\mathcal{P}_{D,\,N}\left(\widehat{a}\,\right)}\;
\frac{\partial\, \mathcal{P}_{D,\,N}\left(\widehat{a}\,\right)}
     {\partial\left(\delta_{\mu\nu}\, \widehat{a}^{\;\nu}_{\;lk}\right)}
-\widehat{\eta}^{\;\mu}_{\;kl} = 0\,,
%\\[2mm]
%\hspace*{-6mm}&&
%\mathcal{P}_{D,\,N} \left(\widehat{a}\,\right) =
%\text{homogeneous polynomial of order $K$}\,,
\end{eqnarray}
%\end{subequations}
where the Pfaffian
$\mathcal{P}_{D,\,N} \left(\widehat{a}\,\right)$ is
a homogeneous polynomial of order $K$ from \eqref{eq:Pfaffian-order-polynomial-K}
and directional indices $\mu,\, \nu$  run over $\{  1,\,  \ldots \,,\,D \}$,
with the repeated indices $\lambda,\nu$ implicitly summed over.
The $N \times N$ matrices $\widehat{\eta}^{\;\mu}$ are also traceless and Hermitian.

%%%%%%%%%%%%%%%%%%%%%%%%%

The algebraic equation \eqref{eq:full-algebraic-equation}
has two types of constants:
the master momenta $\widehat{p}_{k}$
(fixed random numbers from a uniform distribution with a cutoff)
and the master noise  matrices $\widehat{\eta}^{\;\mu}_{\;kl}$
(fixed random numbers from a Gaussian distribution).
Very briefly, the meaning
of these two types of random numbers can be explained as follows.

The dimensionless time $\tau$ is the fictitious
Langevin time of stochastic quantization, with a
Gaussian noise term $\eta$ in the differential equation,
%%(the basic structure of the equation is as follows:
%%$d A(\tau)/d \tau= -\delta S_\text{eff}/\delta A(\tau) +\eta(\tau)$.
\beq
\label{eq:Langevin-diff-eq}
\frac{d A^{\mu}(\tau)}{d \tau}=
-\,\frac{\delta S_\text{eff}(\tau)}{\delta A_{\mu}(\tau)} +\eta^{\mu}(\tau)\,.
\eeq
The $\tau$ evolution drives the system to equilibrium at
$\tau=\tau_\text{eq}$ and the resulting configuration
$A^{\mu}(\tau_\text{eq})$ corresponds to
the master field $\widehat{A}^{\;\mu}$.
For large $N$, the $\tau$-dependence of
the Langevin noise matrices $\eta^{\mu}_{\,k\,l}(\tau)$
can be quenched~\cite{GrossKitazawa1982} by use of the master momenta $\widehat{p}_{k}$
and the same holds for
the corresponding bosonic variables $A^{\mu}_{\,k\,l}(\tau)$.
We then have at $\tau \geq \tau_\text{eq}$\,:
\bsubeqs\label{eq:eta-quenched-A-quenched}
\beqa
\label{eq:eta-quenched}
\eta^{\mu}_{\,k\,l}(\tau) &=&
e^{\displaystyle{i\,\widehat{p}_{k}\,\tau}}
\;\,\widehat{\eta}^{\;\mu}_{\,k\,l}\;\,
e^{\displaystyle{-i\,\widehat{p}_{l}\,\tau}}\,,
\\[2mm]
\label{eq:A-quenched}
A^{\mu}_{\,k\,l}(\tau) &=&
e^{\displaystyle{i\,\widehat{p}_{k}\,\tau}}
\;\,\widehat{a}^{\;\mu}_{\,k\,l}\;\,
e^{\displaystyle{-i\,\widehat{p}_{l}\,\tau}}\,,
\eeqa
\esubeqs
where the matrices $\widehat{\eta}^{\;\mu}$ and $\widehat{a}^{\;\mu}$
on the right-hand sides have no dependence on  $\tau$.
%Note that all master variables and master constants are denoted by a caret.
Using the quenched behavior \eqref{eq:eta-quenched-A-quenched}
with master momenta $\widehat{p}_{k}$,
the Langevin differential equation \eqref{eq:Langevin-diff-eq}
reduces to the algebraic equation \eqref{eq:full-algebraic-equation}.
Most importantly, the random constants $\widehat{p}_{k}$
and $\widehat{\eta}^{\;\mu}_{\,k\,l}$ in \eqref{eq:full-algebraic-equation}
can be \emph{fixed once and for all,} provided $N$ is large enough.
See Refs.~\cite{GreensiteHalpern1983,Klinkhamer2020-master}
for further discussion and references.

%%%%%%%%%%%%%%%%%

Incidentally, we have also considered a simplified version of the
algebraic equation \eqref{eq:full-algebraic-equation},
with all constants $\widehat{p}_{k}$ and $\widehat{\eta}^{\;\mu}_{\;kl}$
set to zero. This  simplified equation corresponds, in fact,
to the stationarity equation of
the effective bosonic action. Some of its solutions, critical points,
will be presented in App.~\ref{app:Nontrivial-critical-points}
(related critical points have been used in
a recent paper~\cite{Steinacker2022}).

Remark that, as the Pfaffian \eqref{eq:calP-calM}
is a $K$-th order homogeneous polynomial
denoted symbolically by $P_{K}[A]$
with $K$ given by \eqref{eq:Pfaffian-order-polynomial-K},
the basic structure of the algebraic
equation \eqref{eq:full-algebraic-equation} is as follows:
\beq
\label{eq:algebraic-equation-structure}
P_{1}^{(\,\widehat{p}\,)}\left[\widehat{a}\,\right]
+P_{3}\left[\widehat{a}\,\right]
+F\,\frac{P_{K-1}\left[\widehat{a}\,\right]}{P_{K}\left[\widehat{a}\,\right]}
+P_{0}^{(\,\widehat{\eta}\,)}\left[\widehat{a}\,\right]=0\,,
\eeq
where only the on/off constant $F$ is shown explicitly
and where the suffixes on $P_{1}$ and $P_{0}$
indicate their respective dependence
on the master momenta $\widehat{p}_{k}$
and the master noise $\widehat{\eta}^{\;\mu}_{kl}$.
If we multiply \eqref{eq:algebraic-equation-structure} by
$P_{K}\left[\widehat{a}\,\right]$, we get a homogeneous polynomial
equation of order $K+3$.

In order to obtain the component equations
[labelled by an index $c$ running over $1,\, \ldots \,,\, (N^2-1)$
and an index $\mu$ running over $1,\, \ldots \,,\, D$],
we matrix multiply \eqref{eq:full-algebraic-equation}
by $t_{c}$, take the trace, and multiply the result
by two.
There are then $D\,g=D\,\big(N^2-1\big)$ coupled algebraic equations
for an equal number of unknowns
$\{\widehat{a}_{1}^{\;1},\, \,\ldots\, ,\, \widehat{a}_{D}^{\;g}\}$.

It appears impossible to obtain a \emph{general} solution of
these algebraic equations. We will
look for solutions of these coupled algebraic equations
with a \emph{specific} realization of the
random  constants $\widehat{p}_{k}$
and $\widehat{\eta}^{\;\mu}_{\;kl}$
[this procedure suffices for large $N$, as mentioned
a few lines below \eqref{eq:A-quenched}].
This is still a formidable problem for large values
of $N$.  At this moment, we are able to get explicit results
only for very small values of $N$.

The  solution of \eqref{eq:full-algebraic-equation}
is given by $D$ traceless matrices $\widehat{a}^{\;\mu}$
of dimension $N \times N$ and the number of real variables reads
\beqa
\label{eq:Nrealvar}
&&
N_\text{real-var} = D\,\big(N^2-1\big)
\times
\begin{cases}
 2 \,,   &    \text{for}\;F=1 \;\text{and}\; N \geq 4 \,,
 \\[2mm]
 1 \,,   &  \text{otherwise}\,,
\end{cases}
\eeqa
where the extra factor of $2$ comes from having a complex
Pfaffian~\cite{NishimuraVernizzi2000-JHEP}, so that the solution
has complex coefficients and non-Hermitian matrices
(see Sec.~\ref{subsec:Num-results-full-alg-eq} for further discussion).
In case complex coefficients are needed, we define
\beqa
\label{eq:complex-coefficients-ahat-mu-c}
&&
\widehat{a}_{\mu}^{\;c} =
\widehat{r}_{\mu}^{\;c}+ i\; \widehat{s}_{\mu}^{\;c}\,,
\eeqa
with a Lie-algebra index $c$ and
real numbers $\widehat{r}_{\mu}^{\;c}$ and $\widehat{s}_{\mu}^{\;c}$.

%%\newpage%%tmp
\subsection{Conceptual and technical remarks}
\label{subsec:Conceptual-remarks}

Before we set out on obtaining solutions of the
algebraic equation \eqref{eq:full-algebraic-equation},
we have two conceptual remarks and one technical remark:
\begin{itemize}
\item%%\vspace*{-2mm}
there are no $\hbar$'s and $G_N$'s in the IIB matrix model \eqref{eq:Z-D-N-F-all-defs},
which is just a ``statistical'' theory,
and we will have to identify the emerging quantum effects
and gravity;
\item%%\vspace*{-2mm}
there is a symmetry argument~\cite{Aoki-etal-review-1999}
for $A_\mu$ having the dimension of length
(that is, not the dimension of inverse length or ``momentum'');
\item%%\vspace*{-2mm}
in our subsequent calculations,
the Pfaffian term in the algebraic equation \eqref{eq:full-algebraic-equation}
for $F=1$ will be obtained \emph{exactly}
(different from the calculations of, e.g., Ref.~\cite{Anagnostopoulos-etal-2020}).
%%\vspace*{-2mm}
\end{itemize}
The meaning of the first and last remarks is clear,
but let us expand on the second remark.

The action $S=S_\text{bos}+S_\text{ferm}$
from \eqref{eq:Z-D-N-F-all-defs}
has two fermionic symmetries (labeled by $i=1,\,2$),   %% of the fields,
which are given by Eqs.~(1.2), (1.4), and (1.5) in
Ref.~\cite{Aoki-etal-review-1999}.
(Incidentally, a useful discussion of the invariance properties of
supersymmetric Yang--Mills gauge theory in $D=3,\,4,\,6,\,10$ spacetime dimensions
can be found in App.~4.~A of Ref.~\cite{GreenSchwarzWitten1987}.)
With two infinitesimal fermionic parameters $\epsilon_{1}$
and $\epsilon_{2}$, the two symmetry generators
$\widetilde{Q}^{(i)}$ have
the following commutation relations~\cite{Aoki-etal-review-1999}:
\beq
\label{eq:Q-Q-relation}
\left[
\bar{\epsilon_{1}}\,\widetilde{Q}^{(i)},\,
\bar{\epsilon_{2}}\,\widetilde{Q}^{(j)}
\right]=
-2\,\bar{\epsilon_{1}}\,\gamma_{\mu}\,\epsilon_{2}\;
\mathcal{P}^{\mu}\,\delta^{i\,j}\,,
\eeq
where $\mathcal{P}^{\mu}$ is the generator of
the following bosonic transformation (similar to a translation):
\beq
\label{eq:translation-transformation}
\delta A_{\mu}= \xi_{\mu}\,\id_{N}\,,
\eeq
for $D$ infinitesimal constants $\xi_{\mu}$ and the
$N\times N$ unit matrix $\id_{N}$. The transformation
\eqref{eq:translation-transformation} obviously leaves
the action $S=S_\text{bos}+S_\text{ferm}$
from \eqref{eq:Z-D-N-F-all-defs} invariant, as $\id_{N}$
trivially commutes with the matrices $A^{\mu}$  and $\Psi_{\alpha}$.

The structure of the commutation relations \eqref{eq:Q-Q-relation}
is precisely that of a ten-dimensional $\mathcal{N}=2$ spacetime supersymmetry
and we may interpret $\mathcal{P}^{\mu}$ as a ``momentum'' with
the dimension of an inverse length.
In turn, this implies that the $A_{\mu}$ matrices
(and, \emph{a fortiori}, their eigenvalues) have the dimension of length.
[Let us clarify the reason for
calling $\mathcal{P}^{\mu}$ a ``momentum'': freely introducing
$\hbar$'s and $c$'s, the Hamiltonian of a supersymmetric theory
is the anticommutator of two supercharges $Q$
(cf. Sec.~5.2.2 of Ref.~\cite{GreenSchwarzWitten1987})
and the energy corresponds to the
0-component of the momentum vector ($E=c\,p^{0}$),
so that the identification of
$\mathcal{P}^{\mu}$ in \eqref{eq:Q-Q-relation}
with a momentum vector makes sense.]

It is, in principle, easy to give the
bosonic matrices $A_{\mu}$ of the IIB matrix model
the dimension of length. For this, we replace
the actions $S_\text{bos}$ and $S_\text{ferm}$
in the exponentials of \eqref{eq:Z-D-N-F}
by $S_\text{bos}/\ell^{4}$ and $S_\text{ferm}/\ell^{4}$,
for a model length scale $\ell$.
But it is very well possible that the
correct version of the IIB matrix model
has no such \emph{ad hoc} length scale. Then, the
matrices $A^{\mu}$ and $\Psi_{\alpha}$ are dimensionless
and the theory has, most likely, conformal symmetry.
In this case, we only notice the \emph{structure} of the
commutation relations~\eqref{eq:Q-Q-relation},
with all quantities being dimensionless
(the matrices $A^{\mu}$ must later get their dimension of length dynamically).
See Sec.~\ref{sec:Conclusion}
for further comments relating to a ``tamed''
big bang~\cite{KlinkhamerVolovik2021} and
subsequent conformal symmetry
breaking~\cite{ColemanWeinberg1973,BirrellDavies1980,Wilczek2012}.

As a last preliminary remark, it may be helpful
to recall the heuristics~\cite{Klinkhamer2021-APPB-review}
for obtaining an emerging classical spacetime
from the bosonic master-field matrices:
\begin{itemize}
  \item
the expectation values
$\left\langle w^{\mu_{1}\,\ldots\,\mu_{m}}
\,\cdots\,w^{\omega_{1}\,\ldots\,\omega_{z}} \right\rangle$
from (\ref{eq:IIB-matrix-model-w-product-vev}), infinitely
many numbers, correspond to a large part of
the information content of the IIB matrix model
(but, of course, not all the information);
  \item\vspace*{-0mm}
that very same information is contained
in the master-field matrices $\widehat{A}^{\,\mu}$,
as these matrices give, to leading order in $N$,
identical numbers from the products
$\widehat{w}^{\,\mu_{1}\,\ldots\,\mu_{m}} \,\ldots\,
\:\widehat{w}^{\,\omega_{1}\,\ldots\,\omega_{z}}$,
where $\widehat{w}$ is the observable $w$
evaluated for the $\widehat{A}^{\,\mu}$, according to \eqref{eq:def-what};
  \item\vspace*{-0mm}
from these master-field matrices $\widehat{A}^{\,\mu}$,
it then appears possible to extract the points and the metric of
an emergent classical spacetime with
a dimensionality less than or equal to $D$
(recall that the original matrices $A^{\,\mu}$
were merely integration variables).
\end{itemize}
One suggestion for an extraction procedure of points and metric has been
presented in Ref.~\cite{Klinkhamer2020-master} and
was reviewed in the appendices of Ref.~\cite{Klinkhamer2021-APPB-review}.

%%\newpage%%tmp
%%\section{Numerical solutions for $\mathbf{(D,\,N)=(10,\,4)}$}
\section{Numerical solutions for (D, N) = (10, 4)}
\label{sec:Numerical-solutions-D10-N4}

\subsection{Diagnostic quantities}
\label{subsec:Diagnostic-quantities}

Approximate numerical solutions of the full
$(D,\,N)=(10,\,4)$ bosonic master-field
equation \eqref{eq:full-algebraic-equation} with $F=1$
have been obtained and will be discussed shortly.
With these approximate numerical solutions,
the complex residues of the 150 component equations
$\text{eq-}\widehat{a}_{\mu}^{\;c}$
are computed (they all vanish for a perfect solution).
The quantity $\text{MaxAbsRes}$ is the maximum of the absolute values
of these residues
and the function $f_\text{penalty}$ is the sum of their squared
absolute values.  Explicitly, these quantities are defined by
\bsubeqs\label{eq:f-penalty-MaxAbsRes}
\beqa
\label{eq:f-penalty}
f_\text{penalty}
&\equiv& \sum_{\mu=1}^{10}\, \sum_{c=1}^{15}\,
\big|\text{eq-}\widehat{a}_{\mu}^{\;c}\big|^{2}\,,
\\[2mm]
\label{eq:MaxAbsRes}
\text{MaxAbsRes} &\equiv&
\text{max}
\Big\{
\big|\text{eq-}\widehat{a}_{1}^{\;1}\big|\,,
\, \ldots \,,\,
\big|\text{eq-}\widehat{a}_{10}^{\;15}\big|
\Big\}\,.
\eeqa
\esubeqs
We get the expression $\text{eq-}\widehat{a}_{\mu}^{\;c}$
from \eqref{eq:full-algebraic-equation}
by performing a matrix multiplication with $t_{c}$,
taking the trace, and multiplying the result by two
[here, the $t_{c}$ are the $SU(4)$ generators
given in  App.~\ref{app:SU4-generators}].

Following the discussion of a previous
paper~\cite{Klinkhamer2021-first-look}, we will first
consider the absolute values of the entries in the
$4\times 4$ matrix $\widehat{a}_{\;1}$, calculate
the average band-diagonal value from 3+4+3 entries and
the average off-band-diagonal value from 3+3 entries,
and get the ratio $R_{1}$ of the average band-diagonal value
over the average off-band-diagonal value.
For the $\mu=2,\,\ldots \,,\, 10$  matrices $\widehat{a}_{\;\mu}$,
we then follow the same procedure and get the ratios
$R_{2},\,\ldots \,,\,R_{10}$.

In order to avoid any confusion,
we give the general definition of the ratio $R$ for an arbitrary
symmetric $4\times 4$ matrix $M$ with nonnegative entries $m[i,\,j]$:
\beq
\label{eq:def-R}
R \equiv
\frac{\left( \sum_{i=1}^{4} m[i,\,i] +2\,\sum_{j=1}^{3} m[j,\,j+1]\right)\Big/10\;}
{\Big( 2\, m[1,\,3] + 2\, m[1,\,4] + 2\, m[2,\,4]\Big)\Big/6}\,,
\eeq
where the symmetry of $M$ has been used to simplify the expression.

%%\newpage%%tmp
\subsection{Numerical results from the full algebraic equation}
\label{subsec:Num-results-full-alg-eq}

In Ref.~\cite{Klinkhamer2021-numsol}, we obtained
approximate numerical solutions of the full
$(D,\,N)=(10,\,4)$ bosonic master-field
equation \eqref{eq:full-algebraic-equation} with $F=1$
and the particular realization (the ``$\kappa$-realization'') of
the pseudorandom numbers as given in App.~\ref{app:Pseudorandom-numbers}.
In that paper, we used a self-made Random-Step (RS) routine
for \textsc{Mathematica} 12.1 (cf. Ref.~\cite{Wolfram1991}),
which could be partially parallelized.

\begin{table}[t]
\vspace*{-0mm}
\begin{center}
\caption{Approximate numerical solutions
of the full $(D,\,N)=(10,\,4)$ bosonic master-field
equation \eqref{eq:full-algebraic-equation} with $F=1$
and pseudorandom constants given
by \eqref{eq:phat-kappa-realization}
and \eqref{eq:etahatcoeff-kappa-realization}.
Shown are the values of the penalty function \eqref{eq:f-penalty},
the  maximum of the absolute values
of the residues \eqref{eq:MaxAbsRes},
and 4 out of 300 coefficients defined
by \eqref{eq:ahat-matrices-kappa-num-sol}.
The numerical calculations used
a self-made Random-Step (RS) routine, which could be partially parallelized.
The results of the first five rows have already been
reported in Ref.~\cite{Klinkhamer2021-numsol} and
the result of the last row is new.
The calculations for these results took about
6 %%frk 10sep2021-24feb2022
months, using a Lenovo T15p notebook
with an Intel Core i7-10750H processor
running \textsc{Mathematica} 12.1.
}
\vspace*{0.5\baselineskip}%%\vspace*{1\baselineskip}
\label{tab-num-sols-full}
\renewcommand{\tabcolsep}{0.70pc}    %% enlarge column spacing
\renewcommand{\arraystretch}{1.25}   %% enlarge line spacing
\begin{tabular}{c|c|c|c}
\hline\hline
$f_\text{penalty}$
&  $\text{MaxAbsRes} $
& $\{ \widehat{r}_{1}^{\;1},\,  \widehat{s}_{1}^{\;1},\,
      \widehat{r}_{10}^{\;15},\,  \widehat{s}_{10}^{\;15}\}$
& $\text{method}$
\\
\hline\hline
$1375.200$   &  $6.81$
%& $\{ -(219/1000), -(279/1000), -(869/2000), -(7/40)  \}$ \\
& $\{  -0.219000, -0.279000, -0.434500, -0.175000 \}$
& $\text{RS}$\\
%& $\text{September 10, 2021}$ \\%$\text{10sep2021}$\\
$422.468$    &  $4.56$
%& $\{ 418183/14200000, -(1686563/4260000), -(765243/2060000), 3532411/123600000 \}$\\
& $\{ 0.0294495, -0.395907, -0.371477, 0.0285794 \}$
& $\text{RS}$\\
%& $\text{September 16, 2021}$ \\%$\text{16sep2021}$\\
$310.932$    &  $3.46$
& $\{ 0.0299337, -0.378790, -0.368679, 0.00749545  \}$
& $\text{RS}$\\
%& $\text{September 18, 2021}$ \\%$\text{18sep2021}$ \\
$209.330$    &  $2.83$
%& $\{ 4/321,-(153/394), -(211/576),5/314  \}$ \\
& $\{ 0.0124611, -0.388325, -0.366319, 0.0159236  \}$
& $\text{RS}$\\
%& $\text{September 22, 2021}$ \\%$\text{22sep2021}$\\
$108.094$    &  $1.83$
%& $\{ -(6/235),-(173/368), -(102/301),23/375  \}$ \\
& $\{ -0.0255319, -0.470109, -0.338870, 0.0613333  \}$
& $\text{RS}$\\
%& $\text{October 7,  2021}$ \\%$\text{07oct2021}$\\
%
%$34.972$     & $0.972$
%& $\{ -0.0671508,-0.584865, -0.350589,-0.0417536  \}$ \\
%& $\text{February 9, 2022}$ \\%$\text{09feb2022}$ \\
%
$33.8776$     & $0.957$
& $\{ -0.0692546,-0.586430, -0.350589,-0.0458760 \}$
& $\text{RS}$\\
%& $\text{February 24, 2022}$ \\%$\text{24feb2022}$ \\
\hline\hline
\end{tabular}
\end{center}
%%\end{table}
\vspace*{4mm}
%%\begin{table}[h]
\vspace*{-0mm}
\begin{center}
\caption{Ratios $R_{\mu}$, as defined by \eqref{eq:def-R},
calculated with the absolute values of the entries
in the transformed  %%diagonalized
matrices \eqref{eq:ahat-matrices-HERM-prime-kappa-num-sol-last-row}
from the approximate numerical solutions
of Table~\ref{tab-num-sols-full}.
The $\mu=1$ matrix has been diagonalized
and the corresponding ratio $R_{1}$ is trivially infinite,
as the off-diagonal matrix elements are zero.
The ten ratios are presented in two batches
of five, in order to facilitate comparison with
Fig.~\ref{fig:Absahatprimef33pt8776}.
}
\vspace*{0.5\baselineskip}%%\vspace*{1\baselineskip}
\label{tab-num-sols-ratios-full}
\renewcommand{\tabcolsep}{1.25pc}    %% enlarge column spacing
\renewcommand{\arraystretch}{1.25}   %% enlarge line spacing
\begin{tabular}{c|c|c}
\hline\hline
$f_\text{penalty}$  &  $\{R_{1},\,R_{2},\, R_{3},\,R_{4},\, R_{5} \}$
&  $\{R_{6},\,R_{7},\, R_{8},\,R_{9},\, R_{10} \}$  \\
\hline\hline
$1375.200$   &
$\{\infty,0.786,1.10,2.01,0.915\}$  &
$\{0.460,0.919,1.10,0.781,0.590\}$
\\
$422.468$    &
$\{\infty,0.933,0.637,1.67,0.398\}$
 &
$\{0.466,2.31,2.59,0.757,0.648\}$
\\
$310.932$    %%frk or $310.931$???
&
$\{\infty,0.971,0.631,1.28,0.506\}$
&
$\{1.11,1.81,1.08,0.552,0.668\}$
\\
$209.330$  &
$\{\infty,1.12,0.651,1.15,0.495\}$
&
$\{1.11,1.22,0.885,0.626,0.727\}$
\\
$108.094$  &
$\{\infty,1.26,0.780,1.13,0.522\}$
&
$\{1.10,1.20,0.934,0.590,0.697\}$
\\
%
%$34.972$   &
%$\{\infty,1.21,0.826,1.17,0.539\}$ &
%$\{0.912,1.31,1.07,0.594,0.618\}$ \\
%
$33.8776$   &
$\{ \infty,1.22,0.822,1.17,0.540\}$ &
$\{ 0.913,1.31,1.07,0.589,0.617\}$
\\
\hline\hline
\end{tabular}
\end{center}
\end{table}

\begin{figure}[t]
%\vspace*{-15mm}
\begin{center}
\hspace*{0mm}  %%
\includegraphics[width=0.95\textwidth]
{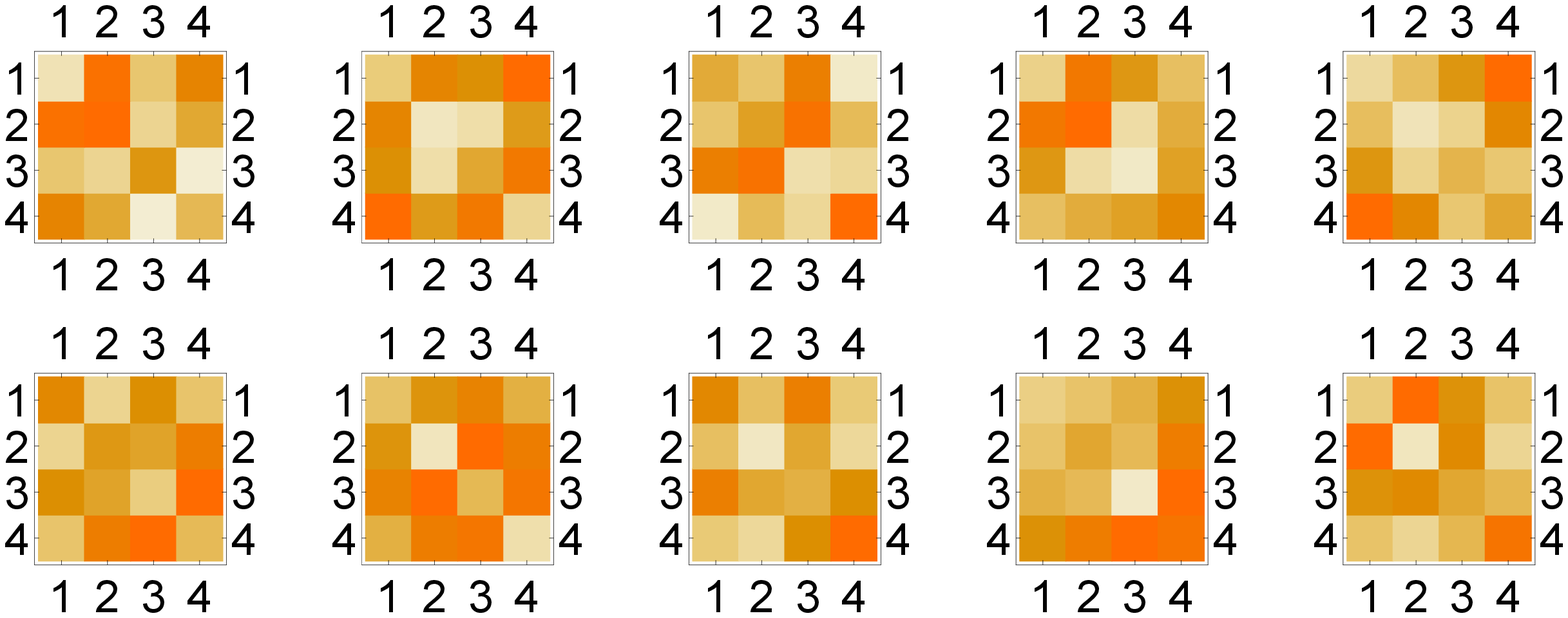}
%%{plotABSahatmu-D10-N4-f33pt8776-24feb2022.eps}
\end{center}\vspace*{-0mm}
\caption{Results from the full $(D,\,N)=(10,\,4)$ bosonic master-field
equation \eqref{eq:full-algebraic-equation} with $F=1$
and pseudorandom constants given by \eqref{eq:phat-kappa-realization}
and \eqref{eq:etahatcoeff-kappa-realization}.
Shown are the density plots of
$\text{Abs}\big[\widehat{a}^{\;\mu}_{\kappa\text{-\underline{num-sol-HERM}}}\big]$
from the approximate solution
with $f_\text{penalty}=33.8776$, as given in Table~\ref{tab-num-sols-full}.
These matrices are defined by \eqref{eq:ahat-matrices-HERM-kappa-num-sol-last-row}.
The panels on the top row are for $\mu=1,\,\ldots\,,5$  and
those on the bottom row for $\mu=6,\,\ldots\,,10$.
%Shown are $\mu=1,\,\ldots\,,5$ on the top row and
%$\mu=6,\,\ldots\,,10$ on the bottom row.
}
\label{fig:Absahatf33pt8776}
\vspace*{0mm}
%%\end{figure}
\vspace*{4mm}
%%\begin{figure}[h]
\begin{center}
\hspace*{0mm}
\includegraphics[width=0.95\textwidth]
{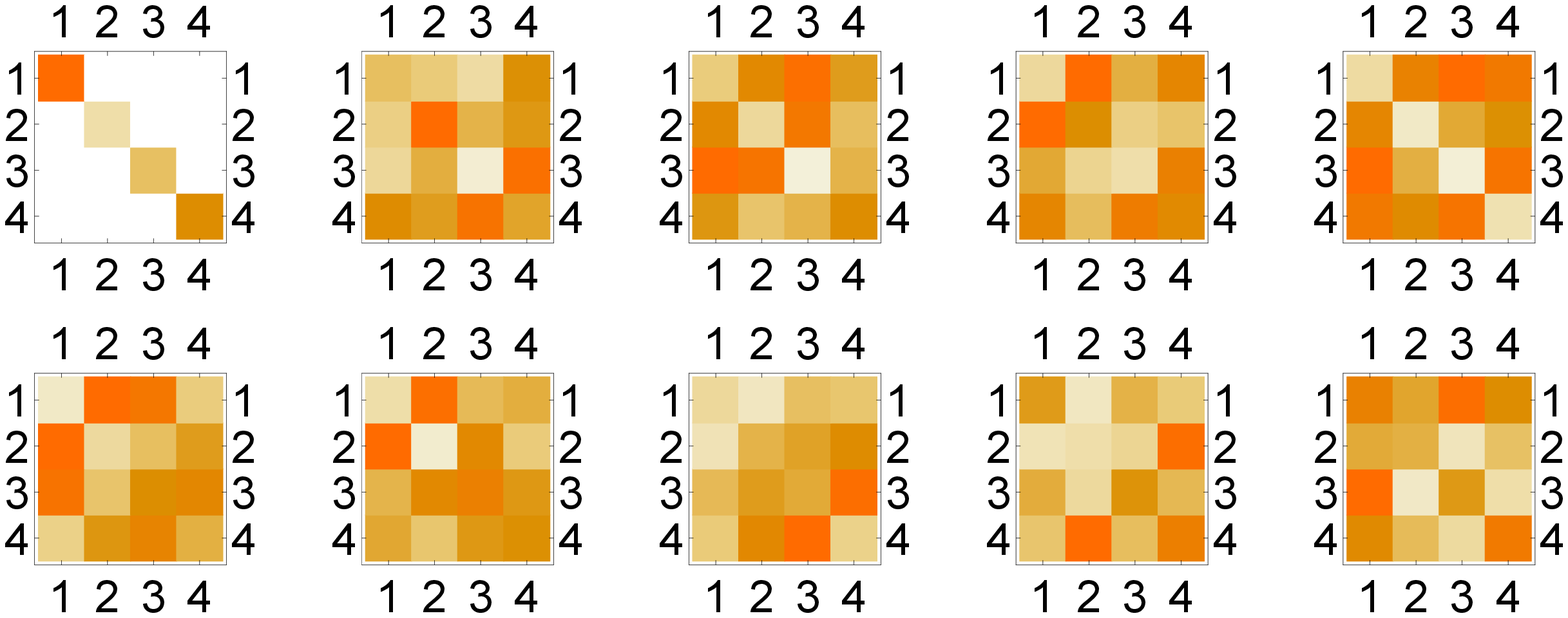}
%%{plotABSahatmuprime-D10-N4-f33pt8776-24feb2022.eps}
\end{center}\vspace*{-0mm}
\caption{Density plots of
$\text{Abs}\big[\widehat{a}^{\;\prime\,\mu}_{\kappa\text{-\underline{num-sol-HERM}}}\big]$
from the matrices of Fig.~\ref{fig:Absahatf33pt8776}
with a change of basis to diagonalize and order one of them ($\mu=1$).
}
\label{fig:Absahatprimef33pt8776}
\vspace*{0mm}
\end{figure}

A short description of this RS routine
is as follows. Given the function value $f_\text{penalty}$  at
a starting point in the 300-dimensional configuration space
(with a trivial Euclidean metric),
the routine calculates the $f_\text{penalty}$  values
at $N_\text{r-p}$  random points
on a sphere centered at this starting point with radius
$r_\text{r-p}$ and moves to the point on the sphere with the
lowest $f_\text{penalty}$  value if that value is better than
the starting-point value,
otherwise the routine takes $N_\text{r-p}$ other random points, etc., etc.
Choosing $N_\text{r-p}$ as an integer multiple of the
number of kernels available, the random points can be
calculated in parallel (we have used, for example,
$N_\text{r-p}=12$ for the $6$ kernels of our processor).
The RS routine is
robust, but requires fine-tuning of the radius of the sphere
and number of random points.

With these approximate numerical solutions, we compute
the complex residues of the 150 component equations
$\text{eq-}\widehat{a}_{\mu}^{\;c}$
(they all vanish for a perfect solution)
and calculate the two diagnostics as defined by \eqref{eq:f-penalty-MaxAbsRes}.
An approximate numerical solution is determined by 300
real numbers and characterized by the corresponding real number
$f_\text{penalty}$. These ten $4\times 4$ matrices are
\beqa
\label{eq:ahat-matrices-kappa-num-sol}
&&
\widehat{a}_{\mu}\,\big|_{\kappa\text{-num-sol}} =
\left(\widehat{r}_{\mu}^{\;c}+ i\; \widehat{s}_{\mu}^{\;c}\right) \,t_{c}\,,
\eeqa
with an implicit sum over the Lie-algebra index $c$ and
real numbers $\widehat{r}_{\mu}^{\;c}$ and $\widehat{s}_{\mu}^{\;c}$.
A selection of  approximate numerical solutions
is shown in Table~\ref{tab-num-sols-full}.
Specifically, we will discuss
the approximate numerical solution from the last row  of
Table~\ref{tab-num-sols-full} with $f_\text{penalty}=33.8776$ and
denote this particular solution by ``$\kappa\text{-\underline{num-sol}}$''.
We, then, have 300 real numbers defining the following matrices:
\beqa
\label{eq:ahat-matrices-kappa-num-sol-last-row}
&&
\widehat{a}^{\,\mu}_{\kappa\text{-\underline{num-sol}}}\,,
\;\;\;\text{for}\;\;\; \mu=1,\,\ldots\,,10\,.
\eeqa
Specifically, the 300 real numbers for
\eqref{eq:ahat-matrices-kappa-num-sol-last-row}
are listed in App.~\ref{app:Coefficients-from-full-algebraic-equation}.

With complex coefficients $\widehat{a}_{\mu}^{\;c}$,
these (approximate) master-field matrices
are no longer Hermitian. The situation is perhaps analogous
to that of pairs of
complex saddle-points appearing for a real problem.
Our interpretation is that these (approximate) master-field matrices
carry \emph{information} both in their Hermitian and anti-Hermitian
parts. In fact, we conjecture that the Hermitian parts of the
master-field matrices (with real eigenvalues)
contain information about the emerging spacetime~\cite{Klinkhamer2020-master}.
But what the information in the anti-Hermitian parts corresponds
to is not clear for the moment (one suggestion has been put forward
in Ref.~\cite{Klinkhamer2021-numsol}).

Consider, therefore, the Hermitian parts
\beqa
\label{eq:ahat-matrices-HERM-kappa-num-sol-last-row}
&&
\widehat{a}^{\,\mu}_{\kappa\text{-num-sol-HERM}}
\equiv
\frac{1}{2}\;
\left[
\widehat{a}^{\,\mu}_{\kappa\text{-num-sol}} +
\left(\widehat{a}^{\,\mu}_{\kappa\text{-num-sol}}\right)^{\dagger}\right]\,.
\eeqa
Calculating the absolute values of these matrix entries
for the $f_\text{penalty}=33.8776$ approximate solution
(denoted ``$\kappa\text{-\underline{num-sol-HERM}}$''),
we observe no obvious band-diagonal structure in Fig.~\ref{fig:Absahatf33pt8776}.

Now, change the basis, in order to
diagonalize and order the $\mu=1$ matrix.
This gives the following transformed matrices denoted by a prime:
\beqa
\label{eq:ahat-matrices-HERM-prime-kappa-num-sol-last-row}
&&
\widehat{a}^{\;\prime\,\mu}_{\kappa\text{-num-sol-HERM}}\,,
\;\;\;\text{for}\;\;\; \mu=1,\,\ldots\,,10\,.
\eeqa
Considering the absolute values of these matrix entries,
there is not yet a strong signal for
a diagonal/band-diagonal structure (density plots are given in
Fig.~\ref{fig:Absahatprimef33pt8776}
for the $f_\text{penalty}=33.8776$ approximate solution).
But let us take a closer look anyway.

\begin{table}[t]
\vspace*{-0mm}
\begin{center}
\caption{Approximate numerical solutions
of the simplified $(D,\,N)=(10,\,4)$ bosonic master-field
equation \eqref{eq:full-algebraic-equation} with $F=0$
and pseudorandom constants given by \eqref{eq:phat-kappa-realization}
and \eqref{eq:etahatcoeff-kappa-realization}.
Two numerical routines have been used:
a self-made Random-Step (RS) routine, which could be partially parallelized,
and the \texttt{NMinimize} (NM) routine of \textsc{Mathematica}
12.1~\cite{Wolfram1991} with the downhill-simplex method
of Nelder and Mead~\cite{NelderMead1965,Press-etal-1992},
which was not parallelized.
The combined numerical calculations took a few days.
}
\vspace*{0.5\baselineskip}%%\vspace*{1\baselineskip}
\label{tab-num-sols-simplified}
\renewcommand{\tabcolsep}{1.25pc}    %% enlarge column spacing
\renewcommand{\arraystretch}{1.25}   %% enlarge line spacing
\begin{tabular}{c|c|c|c}
\hline\hline
$f_\text{penalty}$  &  $\text{MaxAbsRes} $
& $\{ \widehat{r}_{1}^{\;1},\,  \widehat{s}_{1}^{\;1},\,
      \widehat{r}_{10}^{\;15},\,  \widehat{s}_{10}^{\;15}\}$
& $\text{method}$
\\
\hline\hline
$996.965$   &  $7.15$
& $\{ -0.0678381, 0, -0.350114, 0 \}$
& $\text{RS}$  \\
$100.480$  %%frk: extra zero
 &  $2.38$
& $\{ 0.161385,0, -0.340025,0 \}$
& $\text{RS}$  \\
$10.1415$   &  $0.916$
& $\{ -0.124413,0, 0.15631,0 \}$
& $\text{RS}$  \\
$1.21690$  %%frk: extra zero
 &  $0.255$
& $\{ 0.0904698,0, 0.233820,0 \}$  & $\text{RS}$
\\
$0.590364$   &  $0.189$
& $\{ 0.0183176,0, 0.265982,0 \}$
& $\text{RS}$
\\
\hline  %%change method: RS to NM
$0.599656$   &  $0.175$
& $\{ -0.00264037,0, 0.266334,0 \}$  & $\text{NM}$  %%\text{NM-run1}
\\
$0.149264$   &  $0.0879$
& $\{ -0.075562,0, 0.247046,0 \}$ & $\text{NM}$   %%\text{NM-run1}
\\
%$0.0273199$   &  $0.0340$
%& $\{ -0.0617072,0, 0.108870,0 \}$ & $\text{NM}$  %%\text{NM-run1}
%\\
%\hline  %%change NS run
$1.12955 \times 10^{-2}$  %%$0.0112955$
&  $3.28 \times 10^{-2}$  %%$0.0327863$
& $\{ -0.139296,0, 0.128158,0 \}$ & $\text{NM}$  %%\text{NM-run2}
\\
$2.85246  \times 10^{-3}$  %%$0.00285246$
&  $1.40 \times 10^{-2}$  %%$0.0139597$
& $\{ -0.195029,0, 0.111522,0 \}$ & $\text{NM}$  %%\text{NM-run2}
\\
$9.51062  \times 10^{-5}$   %%$0.0000951062$   %%frk: use f-value from original NM calculation
&  $2.77 \times 10^{-3}$   %%$0.00276646$
& $\{ -0.216581,0, 0.122857,0 \}$ & $\text{NM}$  %%\text{NM-run2}
\\
%\hline  %%change NM run
$4.12196\times 10^{-5}$   %%$0.0000412196$   %%frk: use f-value from original NM calculation
&  $1.28 \times 10^{-3}$   %%$0.00128484$
& $\{ -0.218765,0, 0.121856,0 \}$ & $\text{NM}$  %%\text{NM-run3}
\\
$2.32875 \times 10^{-6}$   %%frk: use f-value from original NM calculation
&  $3.75 \times 10^{-4}$  %%$0.000374523$
& $\{ -0.223436,0, 0.121505,0 \}$ & $\text{NM}$  %%\text{NM-run3}
\\
%$3.33614 \times 10^{-7}$   %%frk: use f-value from original NM calculation
%&  $1.44 \times 10^{-4}$   %%$0.000143539$
%& $\{ -0.223364,0, 0.121548,0 \}$ & $\text{NM}$  %%\text{NM-run3}
%\\
%\hline  %%change NM run
$1.33301\times 10^{-7}$  %%frk: use f-value from original NM calculation
&  $7.66\times 10^{-5}$  %%$7.65776\times 10^{-5}$
& $\{ -0.223504,0, 0.121471,0 \}$ & $\text{NM}$  %%\text{NM-run4}
\\
$1.14189\times 10^{-8}$   %%frk: use f-value from original NM calculation
&  $2.57\times 10^{-5}$   %%$2.57012\times 10^{-5}$
& $\{ -0.223699,0, 0.121473,0 \}$ & $\text{NM}$  %%\text{NM-run4}
\\
$2.78202 \times 10^{-9}$   %%frk: use f-value from original NM calculation
&  $1.40\times 10^{-5}$  %%$1.4011\times 10^{-5}$
& $\{ -0.223690,0, 0.121466,0 \}$ & $\text{NM}$  %%\text{NM-run4}
\\
\hline\hline
\end{tabular}
\end{center}
\end{table}

\vspace*{0mm}
\begin{table}[t]
\vspace*{-0mm}
\begin{center}
\caption{Ratios $R_{\mu}$, as defined by \eqref{eq:def-R},
calculated with  the absolute values of the entries in the transformed %%diagonalized
matrices $\widehat{a}_{\kappa\text{-\underline{num-sol-simpl}}}^{\;\prime\,\mu}$
from the approximate numerical solutions
of Table~\ref{tab-num-sols-simplified}.
The ten ratios are presented in two batches
of five, so that the ratios on the last row can be easily
compared with Fig.~\ref{fig:Absahatprime-f3pt02994powerminus9-SIMPLIFIED}.
}
\vspace*{0.5\baselineskip}%%\vspace*{1\baselineskip}
\label{tab-num-sols-ratios-simplified}
\renewcommand{\tabcolsep}{1.0pc}    %% enlarge column spacing
\renewcommand{\arraystretch}{1.25}   %% enlarge line spacing
\begin{tabular}{c|c|c}
\hline\hline
$f_\text{penalty}$  &  $\{R_{1},\,R_{2},\, R_{3},\,R_{4},\, R_{5} \}$
&  $\{R_{6},\,R_{7},\, R_{8},\,R_{9},\, R_{10} \}$  \\
\hline\hline
$996.965$   &
$\{\infty,1.22,0.825,1.17,0.540\}$  &
$\{0.915,1.31,1.07,0.592,0.618\}$
\\
$100.480$   %%frk: extra zero
&
$\{\infty,1.74,1.36,1.08,0.631\}$  &
$\{0.720,0.936,1.42,0.667,0.617\}$
\\
$10.1415$   &
$\{\infty,1.10,1.09,0.718,1.11\}$  &
$\{0.704,0.932,1.21,1.53,0.541\}$
\\
$1.21690$   %%frk: extra zero
&
$\{\infty,0.527,1.76,1.22,0.652\}$  &
$\{0.641,1.00,0.657,1.21,0.350\}$
\\
$0.590364$  &
$\{ \infty,0.627,3.13,1.19,0.793 \}$  &
$\{ 1.01,0.900,0.927,1.09,0.480 \}$
\\
\hline   %%change method: RS to NM
$0.599656$   &
$\{ \infty,0.767,2.65,1.38,0.900 \}$  &
$\{ 1.14,0.876,0.964,1.12,0.497 \}$
\\
$0.149264$   &
$\{ \infty,1.16,2.49,1.36,1.28\}$  &
$\{ 0.949,1.10,1.45,1.44,0.762\}$
\\
%$0.0273199$   &
%$\{ \infty,1.04,2.57,1.20,1.29\}$  &
%$\{ 0.929,1.09,1.39,2.27,0.799\}$
%\\
%\hline  %%change NS run
$1.12955 \times 10^{-2}$  %%$0.0112955$
&  $\{ \infty,1.81,5.27,1.63,1.07 \}$
&  $\{ 1.97,0.92,1.98,3.29,1.09 \}$
\\
$2.85246  \times 10^{-3}$  %%$0.00285246$
&  $\{ \infty,4.00,3.81,1.95,1.14 \}$
&  $\{ 2.59,1.67,1.90,3.49,1.36\}$
\\
$9.51062  \times 10^{-5}$   %%$0.0000951062$ %%frk: use f-value from original NM calculation
&  $\{ \infty,6.07,2.94,2.47,1.70\}$
&  $\{ 3.12,2.86,1.95,3.87,1.63\}$
\\
%\hline  %%change NS run
$4.12196 \times 10^{-5}$  %%$0.0000412196$ %%frk: use f-value from original NM calculation
&  $\{ \infty,6.08,2.91,2.47,1.72 \}$
&  $\{ 3.10,2.95,1.94,3.89,1.61 \}$
\\
$2.32875\times 10^{-6}$  %%frk: use f-value from original NM calculation
&  $\{ \infty,6.06,2.87,2.49,1.75\}$
&  $\{ 3.11,3.07,1.93,3.91,1.60\}$
\\
%$3.33614\times 10^{-7}$  %%frk: use f-value from original NM calculation
%&  $\{ \infty,6.06,2.86,2.49,1.75 \}$
%&  $\{ 3.10,3.07,1.93,3.90,1.60 \}$
%\\
%\hline  %%change NS run
$1.33301 \times 10^{-7}$  %%frk: use f-value from original NM calculation
&  $\{ \infty,6.06,2.86,2.48,1.75 \}$
&  $\{ 3.10,3.08,1.93,3.90,1.60 \}$
\\
$1.14189 \times 10^{-8}$  %%frk: use f-value from original NM calculation
&  $\{ \infty,6.06,2.86,2.49,1.75 \}$
&  $\{ 3.10,3.09,1.93,3.90,1.60 \}$
\\
$2.78202 \times 10^{-9}$  %%frk: use f-value from original NM calculation
&  $\{ \infty,6.06,2.86,2.49,1.75\}$
&  $\{ 3.10,3.09,1.93,3.90,1.60 \}$
\\
\hline\hline
\end{tabular}
\end{center}
\vspace*{0mm}
\end{table}

\begin{figure}[t]
\vspace*{-0mm}
\begin{center}
\hspace*{0mm}
\includegraphics[width=0.95\textwidth]
{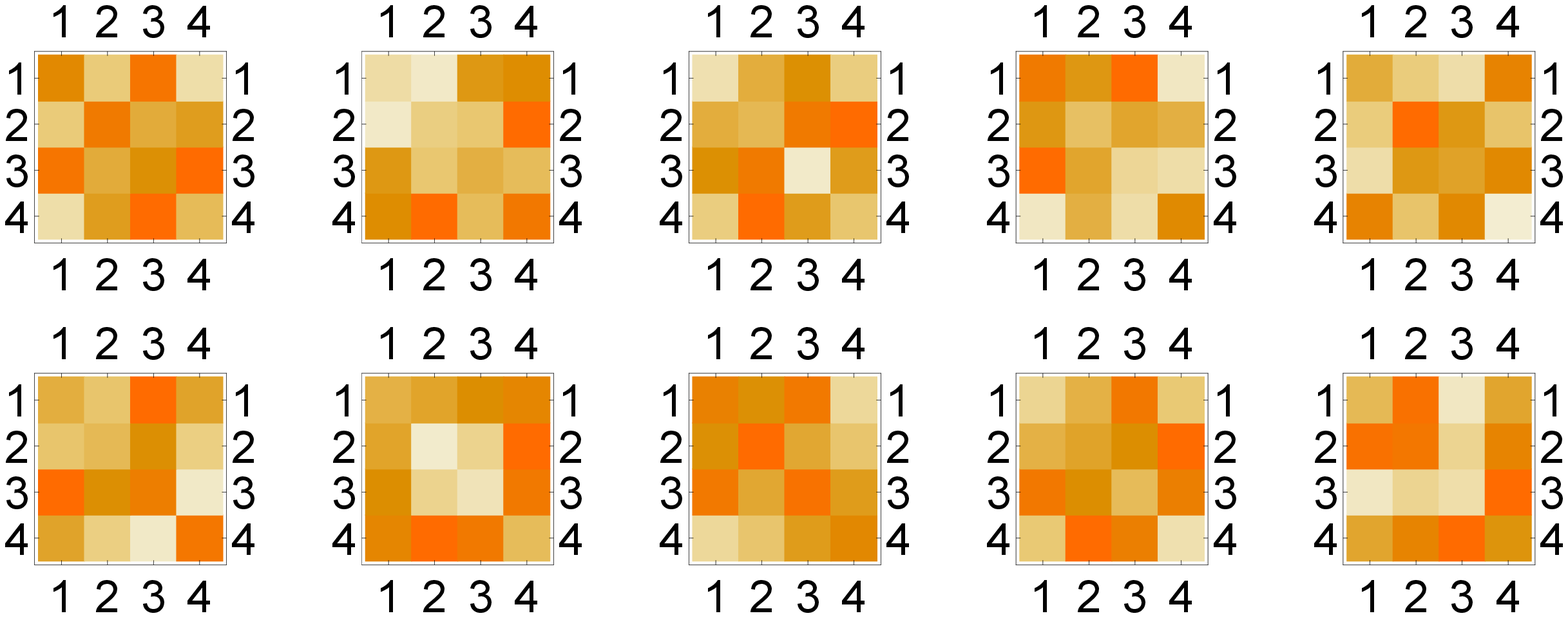}
%%{SIMPLIFIED-SOL-plotABSahatmu-D10-N4-f3pt02994powerminus9-27feb2022.eps}
\end{center}\vspace*{-1mm}
\caption{Results from the simplified $(D,\,N)=(10,\,4)$ bosonic master-field
equation \eqref{eq:full-algebraic-equation} with $F=0$
and the pseudorandom constants given by \eqref{eq:phat-kappa-realization}
and \eqref{eq:etahatcoeff-kappa-realization}.
Shown are the density plots of
$\text{Abs}\big[\widehat{a}^{\;\mu}_{\kappa\text{-\underline{num-sol-simpl}}}\big]$
from the approximate solution
%having $f_\text{penalty}=3.02994 \times 10^{-9}$,
having $f_\text{penalty}=2.78202 \times 10^{-9}$,
as given by Table~\ref{tab-num-sols-simplified}.
The panels on the top row are for $\mu=1,\,\ldots\,,5$  and
those on the bottom row for $\mu=6,\,\ldots\,,10$.
}
\label{fig:Absahat-f3pt02994powerminus9-SIMPLIFIED}
\vspace*{0mm}
%\end{figure}
\vspace*{2mm}
%\begin{figure}[h]
\begin{center}
\hspace*{0mm}
\includegraphics[width=0.95\textwidth]
{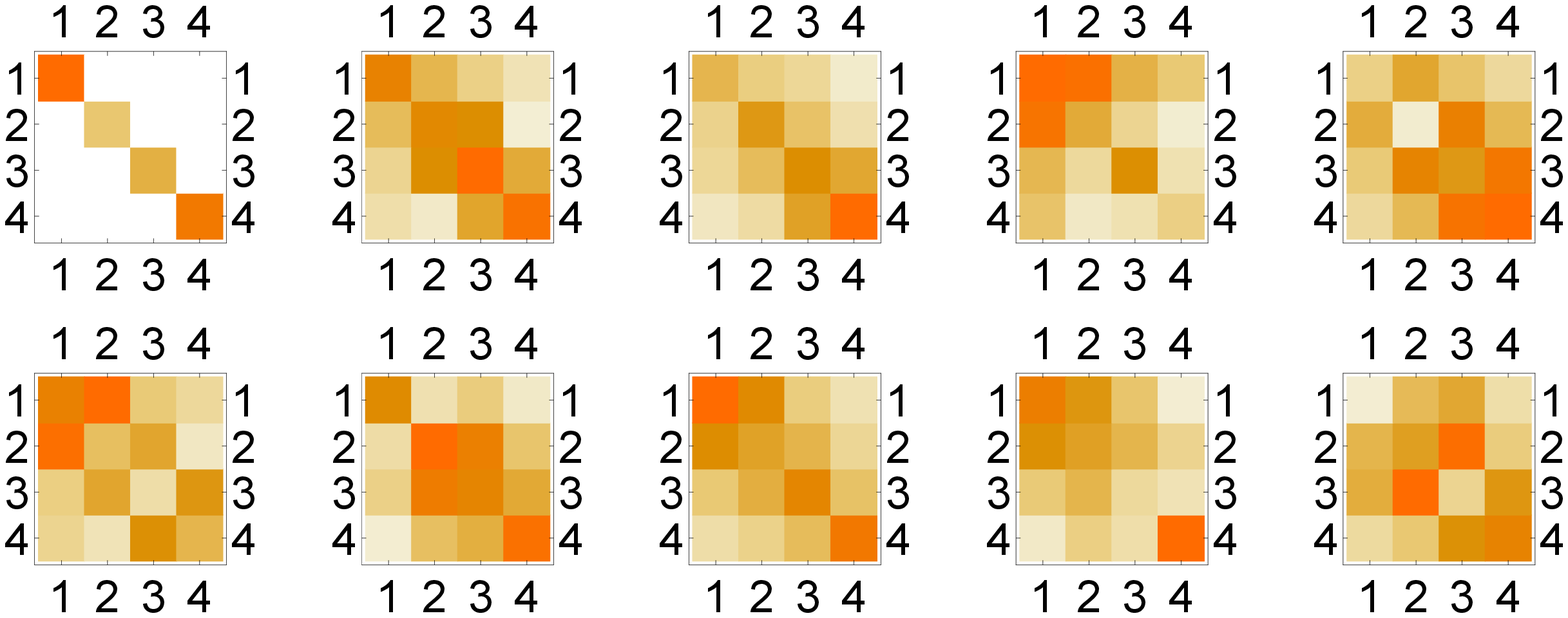}
%%{SIMPLIFIED-SOL-plotABSahatmuprime-D10-N4-f3pt02994powerminus9-27feb2022.eps}
\end{center}\vspace*{-1mm}
\caption{Density plots of
$\text{Abs}\big[\widehat{a}^{\;\prime\,\mu}_{\kappa\text{-\underline{num-sol-simpl}}}\big]$
from the matrices of Fig.~\ref{fig:Absahat-f3pt02994powerminus9-SIMPLIFIED}
with a change of basis to diagonalize and order one of them ($\mu=1$).
}
\label{fig:Absahatprime-f3pt02994powerminus9-SIMPLIFIED}
\vspace*{0mm}
\end{figure}

Using the ratio $R$ defined in \eqref{eq:def-R}, we obtain the
values given in Table~\ref{tab-num-sols-ratios-full}.
On the last row of this table with
$f_\text{penalty}=33.8776$, we see $R_{\mu}$ values
scattered around unity but no clear pattern of band-diagonality.
It may, however, be that this numerical solution
is simply not yet good enough and that the
$f_\text{penalty}$ value needs to be reduced significantly.
At this moment, we are not able to do that for
the full $(D,\,N)=(10,\,4)$ bosonic master-field
equation \eqref{eq:full-algebraic-equation} with $F=1$.
But the simplified equation with $F=0$ does allow us
to push $f_\text{penalty}$ further down, as will be discussed
in Sec.~\ref{subsec:Num-results-simplified-alg-eq} .

%%\newpage%%tmp
\subsection{Numerical results from the simplified algebraic equation}
\label{subsec:Num-results-simplified-alg-eq}

%%first-look-master-field-eq-IIB-mat-mod-v6.tex final for IJMPD
%%11may2021: D2-N4 MATHEMATICA 12.1 on Lenovo Notebook2021

Here, we report on approximate numerical solutions
of the simplified ($F=0$) bosonic master-field
equation \eqref{eq:full-algebraic-equation},
where we start with the same procedure and
self-made Random-Step (RS) routine as has been used for
the full bosonic master-field
equation in Sec.~\ref{subsec:Num-results-full-alg-eq}.
The random-step calculations for the simplified bosonic master-field equation
are much simpler because there are fewer real variables
(now 150 instead of 300) and because the difficult
Pfaffian term is altogether absent in the simplified equation.

Random-step results are presented in Table~\ref{tab-num-sols-simplified}
for penalty-function values of   order $10^{3}$, $10^{2}$,
$10^{1}$, and $1$.
With the computer as described in the caption of
Table~\ref{tab-num-sols-full}, these results for the simplified equation
have taken about 30 hours, which may be compared
with the approximately 6 months for the full-algebraic-equation results
of Table~\ref{tab-num-sols-full}.

But the simplified algebraic equation also allows for the
implementation
of the \texttt{NMinimize} routine of \textsc{Mathematica} 12.1
(cf. Ref.~\cite{Wolfram1991}) with the downhill-simplex method
of Nelder and Mead~\cite{NelderMead1965,Press-etal-1992}.
(It is important to use a trick to force the routine \texttt{NMinimize}
to be purely numerical; see App.~\ref{subapp:Setup-Two-expectation-values} for details.)
These NMinimize results are also shown in Table~\ref{tab-num-sols-simplified}
for penalty-function values of order $10^{-1}$ and down to $10^{-9}$.
\textit{A priori}, there is the  worry that changing the
numerical method gives solutions belonging to different
valleys. But this does not appear to be the case here,
as shown by the two configurations with $f_\text{penalty} \sim 0.6$ obtained
in the different runs (this will be especially clear in
the later Table~\ref{tab-num-sols-ratios-simplified}).

The solution on the last row  of Table~\ref{tab-num-sols-simplified}
with $f_\text{penalty}=2.78202 \times 10^{-9}$
will be denoted  ``$\kappa\text{-\underline{num-sol-simpl}}$''.
We, then, have 150 real numbers defining the following Hermitian matrices:
\beqa
\label{eq:ahat-matrices-kappa-simpl-num-sol-last-row}
&&
\widehat{a}^{\,\mu}_{\kappa\text{-\underline{num-sol-simpl}}}\,,
\;\;\;\text{for}\;\;\; \mu=1,\,\ldots\,,10\,.
\eeqa
A density plot of this solution is shown in
Fig.~\ref{fig:Absahat-f3pt02994powerminus9-SIMPLIFIED}
and a density plot of the  transformed  %%diagonalized
matrices in
Fig.~\ref{fig:Absahatprime-f3pt02994powerminus9-SIMPLIFIED}.

As to the apparent band-diagonality in the density plots of
Fig.~\ref{fig:Absahatprime-f3pt02994powerminus9-SIMPLIFIED},
we can again quantify the analysis by considering
the ratio $R$ defined in \eqref{eq:def-R}.
The obtained values are given in Table~\ref{tab-num-sols-ratios-simplified}.
Apparently, the $R_\mu$ values
have stabilized for $f_\text{penalty} \lesssim 10^{-4}$.
A clear change is observed between the $R_\mu$ values
for $f_\text{penalty} \gtrsim 1$, with an approximately
equal number of values above 1 as below 1,
and the $R_\mu$ values
for $f_\text{penalty} \lesssim 1/100 $,
with all or nearly all values above 1.
This issue will be discussed further in Sec.~\ref{subsec:Discussion-numerical-results}.

%%\newpage%%tmp
\subsection{Discussion of the numerical results}
\label{subsec:Discussion-numerical-results}

In Table~\ref{tab-comparison-num-sols},
we summarize the presently available numerical results
for $N=3$ and $N=4$ at different values of $D$.
There are two tentative conclusions from the results
collected in Table~\ref{tab-comparison-num-sols}:
the strength of the band-diagonality structure appears to be
diminished by the increase of the number of dimensions
[for the $N=4$ and $F=0$ results, from $D=2$ to $D=10$]
and that strength also appears to be
diminished by the inclusion of dynamic fermions
[for the $D=3$ and $N=3$ results, from $F=0$ (without dynamic fermions)
to $F=1$ (with dynamic fermions)].

Note that $f_\text{penalty}$ values of order $10^{-36}$
%in Ref.~\cite{Klinkhamer2021-first-look}
for the $(D,\,N)=(2,\,4)$ simplified algebraic equation
relied on the use of the routine \texttt{FindMinimum}
of \textsc{Mathematica} 12.1~\cite{Wolfram1991},
which is partially algebraic.
That same routine was also used for the $(D,\,N)=(3,\,3)$
results mentioned in Table~\ref{tab-comparison-num-sols}.
It will be hard to achieve these kind of accuracies
with purely numerical methods, but perhaps less radical
values of $f_\text{penalty}$ suffice, as shown
in Table~\ref{tab-num-sols-ratios-simplified}
for the simplified $(D,\,N)=(10,\,4)$ bosonic master-field equation.

For the $(D,\,N)=(10,\,4)$ case, let us now discuss
the possible appearance of a band-diagonal
structure, after one of the master-field matrices has been
diagonalized and ordered.
First, we consider the results of
the simplified $(D,\,N)=(10,\,4)$ bosonic master-field equation
in Table~\ref{tab-num-sols-ratios-simplified}.
Specifically, let us look for a diagonal/band-diagonal pattern
in the density plots of Fig.~\ref{fig:Absahatprime-f3pt02994powerminus9-SIMPLIFIED}
and count how many have a band-diagonal  pattern and how many not
(having instead a ``scattered''  pattern, so that these matrices perhaps
do not contribute to an emerging classical spacetime):
\bsubeqs
\beqa\label{eq:def-n-non-diag}
10 &=&
1\;\Big[\text{diagonal}\Big]+n_\text{b-d}\;\Big[\text{band-diagonal}\Big]
+(9-n_\text{b-d})\;\Big[\text{scattered}\Big]\,,
\\[2mm]
n_\text{b-d} &\in& \big\{0,\,1,\,2,\, \dots , 9 \big\}\,.
\eeqa
\esubeqs
Setting the band-diagonality criterium at $1.50$,
we obtain the following value for the number
of band-diagonal dimensions from the best
approximate numerical solution of Table~\ref{tab-num-sols-ratios-simplified}:
\beq\label{eq:def-n-non-diag-is-9-simpl-alg-eq}
n_\text{b-d}\,
\Big|^{f_\text{penalty}=3.02994  \times 10^{-9},\;R_\mu >1.50}_\text{simpl-alg-eq}
=9\,,
\quad
\text{for}\;\mu = 2,\, \ldots \,,\,9\,.
\eeq
A first conjecture is that, without dynamical fermions,
all nine non-diagonalized  matrices obtain
some form of band-diagonal structure.

Second, we turn to the results of
the full $(D,\,N)=(10,\,4)$ bosonic master-field equation
in Table~\ref{tab-num-sols-ratios-full}.
Taking the results on the last row at face value, we
set the criterium for band-diagonality at
an \emph{ad hoc} value of $1.10$ and get
\beq\label{eq:def-n-non-diag-is-3-full-alg-eq}
n_\text{b-d}\,\Big|^{f_\text{penalty}=33.8776,\;R_\mu >1.10}_\text{full-alg-eq}
\;\stackrel{?}{=}\; 3\,,
\quad
\text{for}\;\mu=2,\,4,\,7\,.
\eeq
Remark that the same three directions are singled out by the
numerical solution with $f_\text{penalty}=108.094$
in Table~\ref{tab-num-sols-ratios-full}.
In order to confirm or disprove the tentative result from
\eqref{eq:def-n-non-diag-is-3-full-alg-eq}, we need to improve
the numerical solution by reducing $f_\text{penalty}$ significantly
(i.e.,  $f_\text{penalty}$  values far below $10$).
Recall that we needed $f_\text{penalty}$ values of order $10^{-4}$ or better
for the $(D,\,N)=(10,\,4)$ simplified-algebraic-equation results
in Table~\ref{tab-num-sols-ratios-simplified}, but perhaps the
full-algebraic-equation results stabilize for somewhat larger $f_\text{penalty}$ values.

\begin{table}[t]
\vspace*{-0mm}
\begin{center}
\caption{Numerical results %%obtained up till now
from the bosonic master-field equation
\eqref{eq:full-algebraic-equation} for $N=3$ and $N=4$,
at different values of $D$ and with ($F=1$)
or without ($F=0$) dynamic fermions.
The hash superscript on ``$\max$'' indicates the restriction to
non-diagonalized directions ($\mu \ne 1$ in the present paper).
The ratio ``$R$'' is defined by \eqref{eq:def-R} for $N=4$ and
\eqref{eq:app-def-Rhat} for $N=3$.
}
\vspace*{0.1\baselineskip}%%\vspace*{1\baselineskip}
\label{tab-comparison-num-sols}
\renewcommand{\tabcolsep}{1.pc}    %% enlarge column spacing
\renewcommand{\arraystretch}{1.25}   %% enlarge line spacing
\begin{tabular}{c|c|c|c|c}
\hline\hline
  $(D,\,N)$   & $\text{algebraic eq. \eqref{eq:full-algebraic-equation}}$
& $f_\text{penalty}$
& $\max^{\#}\left(R_{\mu}\right)$
& $\text{source}$
\\
\hline\hline
$(3,\,3)$ & $\text{simplified}\,(F=0)$ & $\text{O}\left(10^{-69}\right)$
& $\text{O}\left(10\right)$
& $\text{Sec.~4.2\;of\;Ref.~\cite{Klinkhamer2021-sols-D3-N3}}$
\\
$(3,\,3)$ &   $\text{full}\,(F=1)$ & $\text{O}\left(10^{-64}\right)$
& $\text{O}\left(5\right)$
& $\text{Sec.~4.3\;of\;Ref.~\cite{Klinkhamer2021-sols-D3-N3}}$
\\
$(2,\,4)$  & $\text{simplified}\,(F=0)$ & $\text{O}\left(10^{-36}\right)$
& $\text{O}\left(10\right)$
& $\text{Sec.~4.1\;of\;Ref.~\cite{Klinkhamer2021-first-look} }$
\\
$(10,\,4)$  & $\text{simplified}\,(F=0)$ & $\text{O}\left(10^{-9}\right)$
& $\text{O}\left(6\right)$
&  $\text{Sec.~\ref{subsec:Num-results-simplified-alg-eq}\;here}$
\\
$(10,\,4)$  & $\text{full}\,(F=1)$ &$\text{O}\left(10\right)$
& $\text{O}\left(1\right)$
& $\text{Sec.~\ref{subsec:Num-results-full-alg-eq}\;here}$
\\
\hline\hline
\end{tabular}
\end{center}
\vspace*{-4mm}
\end{table}

A second conjecture is that a 3+6 split of the non-diagonalized
matrices (with $n_\text{b-d}=3$)
requires nontrivial fermionic dynamics and possibly supersymmetry.

%%\newpage%%tmp

\section{Conclusion}
\label{sec:Conclusion}%%%\vspace*{-4mm}

In this article, we have explored the hypothesis that a new phase
replaces the Friedmann big bang singularity resulting from our current
theories, general relativity and the
standard model of elementary particle physics.
For such a new phase, we need a new theory which extends
general relativity and the standard model,
and we have used nonperturbative superstring theory in the guise
of the IIB matrix model~\cite{IKKT-1997,Aoki-etal-review-1999}.
The model consists of $N \times N$ traceless Hermitian matrices, with
$10$ bosonic matrices and $16$ fermionic matrices.

The first task at hand is to determine how the IIB matrix model
gives rise to a classical spacetime.
It appears that the required information is encoded in the
master-field matrices of the model~\cite{Klinkhamer2020-master}.
The next task is to calculate these master-field matrices
and to determine what type of spacetime they give.
This is, of course, extremely
difficult. Still, we have been able to determine
the needed master-field equation
and to show the existence of nontrivial solutions
for relatively small values of the matrix size $N$.
However, further progress at large values of $N$
appears to be hard and perhaps new (analytic) insights may be called for.

If the IIB matrix model indeed gives a new phase replacing the big bang,
then we not only need to get an emerging spacetime but also emergent
matter. It may be that matter fields appear as appropriate
perturbations of the master field. These perturbations must have
a very special structure, so that genuine fields appear
in the infrared [for example, a scalar field $\phi(x)$
with the proper spacetime dependence]. This special structure
is clarified by the toy-model calculation of App.~A
in Ref.~\cite{Klinkhamer2020-master}.

Let us now return to an issue briefly mentioned in
Sec.~\ref{subsec:Conceptual-remarks}, namely that of a length scale.
In fact, it may be that the proper IIB matrix model has only dimensionless
matrices $A^{\mu}$ and $\Psi_{\alpha}$
without length scale whatsoever and that the IIB matrix model
produces a phase with conformal symmetry.
This would fit in nicely with the recent suggestion~\cite{KlinkhamerVolovik2021}
of a ``tamed'' big bang as a topological quantum phase transition.
Then, the Friedmann big bang singularity would be replaced by a gapless phase
which evolves into a gapped state corresponding to our present Universe.
The dimensionless
IIB matrix model could give rise to such a gapless phase and perhaps
also to those ``matter'' fields (e.g., a 3-form gauge field)
that produce the $q$-type vacuum variable.
The initial evolution of the universe would then be driven by the motion of $q(t)$ away
from the starting configuration $q=0$, with normal %%ponderable
matter (e.g., the quarks and leptons of the standard model) created from
later $q(t)$ oscillations~\cite{KlinkhamerVolovik2021}.
The subsequent dynamics should also provide a length scale, an energy scale,
and a mass scale by breaking the conformal
symmetry~\cite{ColemanWeinberg1973,%
BirrellDavies1980,Wilczek2012}.

\acknowledgments
It is a pleasure to thank, first,
K.N.~Anagnostopoulos, J.~Nishimura, H.C.~Steinacker, A.~Tsuchiya, and
G.E.~Volovik for useful discussions on various occasions
and, second,
G. Zoupanos, H.C.~Stein\-acker, and K.N.~Anagnostopoulos for organizing the
``Workshop on Quantum Geometry, Field Theory and Gravity''
at the Corfu Summer Institute 2021.

%%\newpage%%tmp
\appendix
%%\section{Large-$\mathbf{N}$ factorization}
\section{Large-N factorization}
\label{app:Large-N-factorization}

\subsection{Setup: Two expectation values}
\label{subapp:Setup-Two-expectation-values}

We consider the following two bosonic observables:
\bsubeqs\label{eq:def-w11-w11+11}
\beqa
w_{11} &\equiv&
\frac{1}{N}\;
\text{Tr}\,\big( A^{1}\, A^{1}\big)\,,
\\[1mm]
w_{11+11}&\equiv&
\left[ \frac{1}{N}\;\text{Tr}\,\big( A^{1}\, A^{1}\big)\right]\,
\left[ \frac{1}{N}\;\text{Tr}\,\big( A^{1}\, A^{1}\big)\right]\,,
\eeqa
\esubeqs
and wish to calculate their expectation values
(with short-hand notations $W_{11}$ and $W_{11+11}$),%
\bsubeqs\label{eq:def-W11-W11+11-Z}
\beqa
\label{eq:def-W11}
W_{11} &\equiv&
\langle w_{11} \rangle =
\frac{1}{Z}\;\int[DA]\,w_{11}\,,
\\[1mm]
\label{eq:def-W11+11}
W_{11+11} &\equiv&
\langle w_{11+11} \rangle =
\frac{1}{Z}\;\int[DA]\,w_{11+11}\,,
\\[1mm]
Z &=& \int dA\, \left( \mathcal{P}_{D,\,N} \right)^{F}\,
e^{\displaystyle{-\,S_{\text{bos}}}}\,    %%e^{-S_\text{bos}}
\equiv
\int[DA]\,,
\eeqa
\esubeqs
where
the measure $dA$ is defined by \eqref{eq:Z-D-N-F},
the bosonic action $S_\text{bos}$ by \eqref{eq:Sbos-IIB},
and the Pfaffian $\mathcal{P}_{D,\,N}$ by \eqref{eq:calP-calM}.
The discrete parameter $F$ takes values in $\{ 0,\, 1 \}$.
%The discrete parameter $F$ allows us to include ($F=1$)
%or exclude ($F=0$) the fermion dynamics.

There are then three multi-dimensional integrals to perform
in \eqref{eq:def-W11-W11+11-Z}, each having
\mbox{$D\,(N^2-1)$} dimensions. For these integrals
(assumed to be convergent, see below), we use
the \texttt{NIntegrate} routine of \textsc{Mathematica} 12.1
(cf. Ref.~\cite{Wolfram1991}) with the Adapted-Monte-Carlo method
and split the calculation into many calculations by taking successive shells.
It is important to use a trick to force the routine \texttt{NIntegrate}
to avoid any algebraic steps and to stay purely numerical.
The trick can be explained by a simple example:
\begin{verbatim}
f[y_] := NIntegrate[2*z, {z, 0, y}];
result = NIntegrate[If[x == 0 || x != 0, f[x]], {x, 0, 1}];
\end{verbatim}
%%\texttt{}
where the first line defines a quadratic function $f$ that only works properly
for a numerical variable and where the ``If'' conditional on the second line forces
the integration variable $x$ to be numeric.

With the obtained expectation values $W_{11}$ and $W_{11+11}$, we
determine the following quantity:
\beq
\label{eq:def-DeltaW11+11}
\Delta W_{11+11} \equiv W_{11+11} -\left(W_{11}\right)^2 \,,
\eeq
which tests for large-$N$ factorization
\eqref{eq:IIB-matrix-model-w-product-vev-factorized}.

Our main results are for the $(D,\,N)=(4,\,4)$ bosonic model
with $F=0$. These results have been extended in two ``directions'':
larger matrices ($N=6$) and the inclusion of the fermion dynamics ($F=1$).
For the two bosonic cases ($F=0$ and $N=4,\,6$),
the three integrals of \eqref{eq:def-W11-W11+11-Z}
have been proven to be convergent~\cite{AustingWheater2001}.
For the supersymmetric case ($F=1$ and $N=4$), the two integrals
\eqref{eq:def-W11} and \eqref{eq:def-W11+11}
have not been proven to be convergent~\cite{AustingWheater2001} but they
may still be.

\begin{table}[t]
\vspace*{-0mm}
\begin{center}
\caption{Numerical results from the bosonic  ($F=0$)
and supersymmetric ($F=1$) models for $D=4$. The relevant
quantities have been defined in \eqref{eq:def-w11-w11+11}
and \eqref{eq:def-W11-W11+11-Z}.
All numerical results are only approximative,
the least reliable being those of the fourth column with $F=1$.
}
\vspace*{0.5\baselineskip}%%\vspace*{1\baselineskip}
\label{tab-num-results-factorization}
\renewcommand{\tabcolsep}{0.65pc}    %% enlarge column spacing
\renewcommand{\arraystretch}{1.25}   %% enlarge line spacing
\begin{tabular}{c|c|c|c}
\hline\hline
&   $\{D,\,N,\,F\}=\{4,\,4,\,0\}$
&   $\{D,\,N,\,F\}=\{4,\,6,\,0\}$
&   $\{D,\,N,\,F\}=\{4,\,4,\,1\}$
\\
\hline\hline
$Z$               & $4.70\times 10^{12}$ & $6.07\times 10^{21}$
& $5.47 \times 10^{18}$ \\
$\int [DA]\,w_{11}$    & $2.71\times 10^{12}$ & $4.43\times 10^{21}$
& $4.19 \times 10^{18}$ \\
$\int [DA]\,w_{11+11}$ & $1.62\times 10^{12}$ & $3.28\times 10^{21}$
& $2.89 \times 10^{18}$ \\
$W_{11}$          & $0.577$              & $0.730$
& $0.767$ \\
$W_{11+11}$       & $0.345$              & $0.540$
& $0.529$ \\
\hline\hline
\end{tabular}
\end{center}
\vspace*{0mm}
\end{table}

%%\newpage%%tmp
%%\subsection{Bosonic model for $(D,\,N)=(4,\,4)$ and $(D,\,N)=(4,\,6)$}
\subsection{Bosonic model for (D, N) = (4, 4) and (D, N) = (4, 6)}
\label{subapp:Bosonic-model-D4-N4}

The simplest model we have studied is the
$(D,\,N)=(4,\,4)$ bosonic model, where the
$SU(4)$ generators are given in App.~\ref{app:SU4-generators}.
Preliminary numerical results for the
quantities defined in Sec.~\ref{subapp:Setup-Two-expectation-values}
appear in the second column of Table~\ref{tab-num-results-factorization}.
From these results, we get for the quantity defined by \eqref{eq:def-DeltaW11+11}
the following numerical value:
\beq
\label{eq:DeltaW11+11-N4-F0}
\Delta W_{11+11}^{\,(D=4,\,N=4,\,F=0)} \approx 0.0121\,,
\eeq
which shows the approximate equality of $\langle w_{11+11} \rangle$
and $\langle w_{11} \rangle\,\langle w_{11} \rangle$ at the $4\,\%$ level.

We can extend the model by going to a larger matrix size, $N=6$.
The $SU(6)$ generators are then given by the 15 generators from
App.~\ref{app:SU4-generators} embedded in $6\times 6$ matrices
and 20 additional generators.
From the results given in the third column of
Table~\ref{tab-num-results-factorization}, we get
\beq
\label{eq:DeltaW11+11-N6-F0}
\Delta W_{11+11}^{\,(D=4,\,N=6,\,F=0)} \approx 0.0071\,,
\eeq
which shows a cancellation at the $1\,\%$ level.

%%\newpage%%tmp
%%\subsection{Supersymmetric model for $(D,\,N)=(4,\,4)$}
\subsection{Supersymmetric model for (D, N) = (4, 4)}
\label{subapp:Supersymmetric-model-D4-N4}

For $D=4$, the Pfaffian can be written as
the determinant of a $2\,(N^2-1) \times 2\,(N^2-1)$
complex matrix~\cite{KrauthNicolaiStaudacher1998}:
\beq
\label{eq:Pfaffian-D4-as-det}
\mathcal{P}_{4,\,N}
=\det
\left(
  \begin{array}{cc}
\mathbf{X}_{4}+i\,\mathbf{X}_{3}  &\;\; i\,\mathbf{X}_{2}+\mathbf{X}_{1} \\
i\,\mathbf{X}_{2}-\mathbf{X}_{1}  &\;\; \mathbf{X}_{4}-i\,\mathbf{X}_{3} \\
  \end{array}
\right)\,,
\eeq
in terms of matrices $ \mathbf{X}_{\mu}$
in the adjoint representation of $SU(N)$,
\bsubeqs\label{eq:adjoint-representation-structure constants}
\beqa
\label{eq:adjoint-representation}
\left( \mathbf{X}_{\mu} \right)^{ab}&=&f^{abc}\,A_{\mu}^{c}\,,
\\[2mm]
\label{eq:structure-constants}
f^{abc} &=& -2\,i\,\text{Tr}\,\big( A^{a}\, \big[A^{b},\, A^{c}\big]\big)\,,
\eeqa
\esubeqs
where the $f^{abc}$ are the $SU(N)$ structure constants.
The Pfaffian for the $(D,\,N)=(4,\,4)$ case is then given by
the determinant of a $30\times 30$ complex matrix.
This determinant cannot be
calculated algebraically but can be evaluated numerically
(for this reason, we must force the routine \texttt{NIntegrate}
to stay purely numerical, as discussed in
App.~\ref{subapp:Setup-Two-expectation-values}).

From the results given in the fourth column of
Table~\ref{tab-num-results-factorization}, we get
\beq
\label{eq:DeltaW11+11-N4-F1}
\Delta W_{11+11}^{\,(D=4,\,N=4,\,F=1)} \approx -0.059\,,
\eeq
which shows a cancellation at the $11\,\%$ level.
Incidentally, there is no problem
with obtaining a negative number, which may anyway
still change to a positive number with increasing accuracy.

%%\newpage%%tmp
\subsection{Discussion of the factorization results}
\label{subapp:Discussion}

It is instructive to compare our results
\eqref{eq:DeltaW11+11-N4-F0} and \eqref{eq:DeltaW11+11-N6-F0}
for the $D=4$ bosonic model and to write them as follows:
\bsubeqs\label{eq:DeltaW11+11-N4-N6-F0-1/N2}
\beqa
\label{eq:DeltaW11+11-N4-F0-1/N2}
\Delta W_{11+11}^{\,(D=4,\,N=4,\,F=0)}
&\approx&  %%0.0121\,,
\frac{0.194}{16}\,,
\\[2mm]
\label{eq:DeltaW11+11-N6-F0-1/N2}
\Delta W_{11+11}^{\,(D=4,\,N=6,\,F=0)}
&\approx&  %%0.0071\,,
\frac{0.256}{36}\,.
\eeqa
\esubeqs
This shows that we have mild evidence for an $1/N^2$
behavior of the remnant term, with an $1/N^2$ coefficient
for $N=4$ and $N=6$
that has the same sign (plus) and is of the same order of magnitude
(one tenth).
For the moment, we do not have similar results
for the $D=4$ supersymmetric model.

As mentioned in Sec.~\ref{subsec:Bosonic-master-field},
Ref.~\cite{Ambjorn-etal2000} has already obtained
extensive numerical results
in support of large-$N$ factorization, also for the
four-dimensional version of the IIB matrix model.
These numerical results, which could even reach
a matrix size of $N=48$, considered Wilson-loop-type
and Polyakov-line-type observables,
\bsubeqs\label{eq:def-Wilson-loop-Polyakov-line}
\beqa
o_\text{loop}(k) &\equiv&
\frac{1}{N}\;
\text{Tr}\,\left(
%%e^{i\,k\,A^{1}}\,e^{i\,k\,A^{2}}\,e^{-i\,k\,A^{1}}\,e^{-i\,k\,A^{2}}\,
e^{\displaystyle{i\,k\,A^{1}}}\,e^{\displaystyle{i\,k\,A^{2}}}\,
e^{\displaystyle{-i\,k\,A^{1}}}\,e^{\displaystyle{-i\,k\,A^{2}}}\,
\right)\,,
\\[1mm]
o_\text{line}(k)&\equiv&
\frac{1}{N}\;
\text{Tr}\,\left(e^{\displaystyle{i\,k\,A^{1}}}\right)\,,
\eeqa
\esubeqs
for a real dimensionless parameter $k$.
In the above observables, there appear the $SU(N)$ group elements
$e^{i\,k\,A^{\mu}}$, as may be appropriate for the
study of a Yang--Mills gauge theory with, for example,
the ``area-law'' behavior of $\langle o_\text{loop} \rangle$
as shown in Fig.~4 of Ref.~\cite{Ambjorn-etal2000}.
We have, instead, considered observables
\eqref{eq:def-w11-w11+11}
directly made out products of the Lie-algebra elements $A^{\mu}$,
as may be appropriate for the study of the emergence of spacetime.
In any case, the results for both types
of observables, even with a smaller number of dimensions ($D=4$)
than needed ($D=10$), confirm the property of large-$N$
factorization \eqref{eq:IIB-matrix-model-w-product-vev-factorized},
which is crucial for the existence of a large-$N$ bosonic master field as
discussed in Sec.~\ref{sec:Bosonic-master-field-and-master-field-equation}.

%%\newpage%%tmp
%\section{$\mathbf{SU(4)}$ generators}
\section{SU(4) generators}
\label{app:SU4-generators}

%num-sol-master-eq-v5.tex final version  APPB-53-1-A5(2022)

We now give the explicit realization used for the
$SU(4)$ generators:
%\bsubeqs
%\beqa
\begin{align}
\label{eq:SU4-generators}
\hspace*{+1.0mm}
t_{1} &=
\frac{1}{2}\,\left(
\begin{array}{cccc}
 0 & 1 & 0 & 0 \\
 1 & 0 & 0 & 0 \\
 0 & 0 & 0 & 0 \\
 0 & 0 & 0 & 0 \\
\end{array}
\right)\,,
\;\;%%\quad
&t_{2} &=
\frac{1}{2}\,
\left(
\begin{array}{cccc}
 0 & i & 0 & 0 \\
 -i & 0 & 0 & 0 \\
 0 & 0 & 0 & 0 \\
 0 & 0 & 0 & 0 \\
\end{array}
\right)\,,
\;\;%%\quad
&t_{3} &=
\frac{1}{2}\,\left(
\begin{array}{cccc}
 0 & 0 & 1 & 0 \\
 0 & 0 & 0 & 0 \\
 1 & 0 & 0 & 0 \\
 0 & 0 & 0 & 0 \\
\end{array}
\right)\,,
%%\eeqa
\nonumber\\[1.5mm]
%%\beqa
\hspace*{+1.0mm}
t_{4} &=
\frac{1}{2}\,\left(
\begin{array}{cccc}
 0 & 0 & i & 0 \\
 0 & 0 & 0 & 0 \\
 -i & 0 & 0 & 0 \\
 0 & 0 & 0 & 0 \\
\end{array}
\right)\,,
\;\;%%\quad
&t_{5} &=
\frac{1}{2}\,\left(
\begin{array}{cccc}
 0 & 0 & 0 & 1 \\
 0 & 0 & 0 & 0 \\
 0 & 0 & 0 & 0 \\
 1 & 0 & 0 & 0 \\
\end{array}
\right)\,,
\;\;%%\quad
&t_{6} &=
\frac{1}{2}\,\left(
\begin{array}{cccc}
 0 & 0 & 0 & i \\
 0 & 0 & 0 & 0 \\
 0 & 0 & 0 & 0 \\
 -i & 0 & 0 & 0 \\
\end{array}
\right)\,,
%%\eeqa
\nonumber\\[1.5mm]
%%\beqa
\hspace*{+1.0mm}
t_{7} &=
\frac{1}{2}\,\left(
\begin{array}{cccc}
 0 & 0 & 0 & 0 \\
 0 & 0 & 1 & 0 \\
 0 & 1 & 0 & 0 \\
 0 & 0 & 0 & 0 \\
\end{array}
\right)\,,
\;\;%%\quad
&t_{8} &=
\frac{1}{2}\,\left(
\begin{array}{cccc}
 0 & 0 & 0 & 0 \\
 0 & 0 & i & 0 \\
 0 & -i & 0 & 0 \\
 0 & 0 & 0 & 0 \\
\end{array}
\right)\,,
\;\;%%\quad
&t_{9} &=
\left(
\begin{array}{cccc}
 0 & 0 & 0 & 0 \\
 0 & 0 & 0 & 1 \\
 0 & 0 & 0 & 0 \\
 0 & 1 & 0 & 0 \\
\end{array}
\right)\,,
%%\eeqa
\nonumber\\[1.5mm]
%%\beqa
\hspace*{+1.0mm}
t_{10} &=
\frac{1}{2}\,\left(
\begin{array}{cccc}
 0 & 0 & 0 & 0 \\
 0 & 0 & 0 & i \\
 0 & 0 & 0 & 0 \\
 0 & -i & 0 & 0 \\
\end{array}
\right)\,,
\;\;%%\quad
&t_{11} &=
\frac{1}{2}\,\left(
\begin{array}{cccc}
 0 & 0 & 0 & 0 \\
 0 & 0 & 0 & 0 \\
 0 & 0 & 0 & 1 \\
 0 & 0 & 1 & 0 \\
\end{array}
\right)\,,
\;\;%%\quad
&t_{12} &=
\frac{1}{2}\,\left(
\begin{array}{cccc}
 0 & 0 & 0 & 0 \\
 0 & 0 & 0 & 0 \\
 0 & 0 & 0 & i \\
 0 & 0 & -i & 0 \\
\end{array}
\right)\,,
%%eeqa
\nonumber\\[1.5mm]
%%\beqa
\hspace*{+1.0mm}
t_{13} &=
\frac{1}{2}\,\left(
\begin{array}{cccc}
 1 & 0 & 0 & 0 \\
 0 & -1 & 0 & 0 \\
 0 & 0 & 0 & 0 \\
 0 & 0 & 0 & 0 \\
\end{array}
\right)\,,
\;\;%%\quad
&t_{14} &=
\frac{1}{2}\,\left(
\begin{array}{cccc}
 0 & 0 & 0 & 0 \\
 0 & 0 & 0 & 0 \\
 0 & 0 & 1 & 0 \\
 0 & 0 & 0 & -1 \\
\end{array}
\right)\,,
%%%\\[2mm]
\;\;%%\quad
&t_{15} &=
\frac{1}{2 \sqrt{2}}\,\left(
\begin{array}{cccc}
 1 & 0 & 0 & 0 \\
 0 & 1 & 0 & 0 \\
 0 & 0 & -1 & 0 \\
 0 & 0 & 0 & -1 \\
\end{array}
\right)\,.
\end{align}
%\eeqa
%\esubeqs
%
%\begin{align}
%x&=y           &  w &=z              &  a&=b+c \nonumber\\
%2x&=-y         &  3w&=\frac{1}{2}z   &  a&=b   \nonumber\\
%-4 + 5x&=2+y   &  w+2&=-1+w          &  &
%\end{align}
%
These generators obey the trace condition \eqref{eq:trace-t-t}.

%%\newpage%%tmp
%%\section{Pseudorandom numbers for $\mathbf{(D,\,N)=(10,\,4)}$}
\section{Pseudorandom numbers for (D, N) = (10, 4)}
\label{app:Pseudorandom-numbers}

%num-sol-master-eq-v5.tex final version  APPB-53-1-A5(2022)

In this appendix, we give the
particular realization (the ``$\kappa$-realization'') of
the pseudorandom numbers used for the
approximate numerical solutions of Sec.~\ref{sec:Numerical-solutions-D10-N4}.

Specifically, we take the following 4 real pseudorandom numbers $\widehat{p}_{\mu}$
for the master momenta:
\beqa
\label{eq:phat-kappa-realization}
\hspace*{-5.000mm}&&
\widehat{p}_\text{$\kappa$-realization}
=
\left\{{-\frac{111}{250},\,\frac{19}{200},\,
       -\frac{63}{200},\,\frac{189}{1000}}\right\}\,, %%\scriptstyle
\eeqa
and the following 150 real pseudorandom numbers $\widehat{\eta}^{\,\mu}_{c}$
entering the Hermitian master-noise matrices:%
\bsubeqs\label{eq:etahatcoeff-kappa-realization}
\beqa
\hspace*{-2.000mm}
\left.
\left\{\widehat{\eta}^{\,1}_{1}\,, \ldots \,,\, \widehat{\eta}^{\,1}_{15}\right\}
\right|_\text{$\kappa$-realization}&=&
%%1
\Big\{\frac{1}{1000},\,-\frac{9}{20},\,\frac{353}{1000},\,
\frac{12}{25},\,-\frac{987}{1000},\,-\frac{51}{200},\,
\frac{1}{4},\,-\frac{63}{1000},\,
\nonumber\\[1mm] \hspace*{-2.000mm}&&
-\frac{131}{200},\,-\frac{367}{1000},\,-\frac{169}{200},\,
\frac{171}{250},\,-\frac{151}{250},\,-\frac{369}{500},\,
\frac{593}{1000}\Big\}\,,
%%\eeqa
\\[4.00mm]
%%\beqa
\hspace*{-2.000mm}
\left. \left\{\widehat{\eta}^{\,2}_{1}\,, \ldots \,,\, \widehat{\eta}^{\,2}_{15}\right\} \right|_\text{$\kappa$-realization}&=&
%%2
\Big\{-\frac{153}{250},\,-\frac{47}{250},\,\frac{897}{1000},\,
-\frac{61}{500},\,\frac{269}{1000},\,-\frac{237}{1000},\,
\frac{1}{125},\,-\frac{13}{100},\,
\nonumber\\[1mm] \hspace*{-2.000mm}&&
-\frac{103}{200},\,\frac{367}{500},\,\frac{1}{10},\,
-\frac{71}{1000},\,\frac{69}{1000},\,-\frac{123}{125},\,
-\frac{17}{25}\Big\}\,,
\eeqa
%%\\[4.00mm]
\beqa
\hspace*{-2.000mm}
\left. \left\{\widehat{\eta}^{\,3}_{1}\,, \ldots \,,\, \widehat{\eta}^{\,3}_{15}\right\} \right|_\text{$\kappa$-realization}&=&
%%3
\Big\{\frac{7}{100},\,\frac{431}{1000},\,\frac{17}{20},\,
-\frac{59}{125},\,-\frac{437}{1000},\,\frac{69}{500},\,
-\frac{991}{1000},\,-\frac{49}{125},\,
\nonumber\\[1mm] \hspace*{-2.000mm}&&
\frac{83}{500},\,-\frac{279}{1000},\,\frac{49}{125},\,
\frac{121}{250},\,\frac{313}{500},\,\frac{871}{1000},\,
\frac{7}{125}\Big\}\,,
%%\eeqa
\\[4.00mm]
%%\beqa
\hspace*{-2.000mm}
\left. \left\{\widehat{\eta}^{\,4}_{1}\,, \ldots \,,\, \widehat{\eta}^{\,4}_{15}\right\} \right|_\text{$\kappa$-realization}&=&
%%4
\Big\{-\frac{11}{125},\,-\frac{469}{1000},\,-\frac{439}{500},\,
-\frac{483}{1000},\,-\frac{11}{25},\,\frac{41}{200},\,
\frac{409}{500},\,-\frac{343}{1000},\,
\nonumber\\[1mm] \hspace*{-2.000mm}&&
\frac{3}{8},\,-\frac{407}{500},\,-\frac{141}{250},\,
-\frac{31}{1000},\,\frac{293}{500},\,-\frac{12}{125},\,
\frac{429}{1000}\Big\}\,,
\eeqa
%%\\[4.00mm]
\beqa
\hspace*{-2.000mm}
\left. \left\{\widehat{\eta}^{\,5}_{1}\,, \ldots \,,\, \widehat{\eta}^{\,5}_{15}\right\} \right|_\text{$\kappa$-realization}&=&
%%5
\Big\{\frac{74}{125},\,\frac{249}{500},\,-\frac{511}{1000},\,
-\frac{43}{250},\,-\frac{129}{1000},\,\frac{19}{100},\,
\frac{49}{200},\,-\frac{463}{500},\,
\nonumber\\[1mm] \hspace*{-2.000mm}&&
\frac{23}{100},\,\frac{433}{1000},\,\frac{821}{1000},\,
\frac{199}{1000},\,-\frac{96}{125},\,-\frac{92}{125},\,
\frac{183}{250}\Big\}\,,
%%\eeqa
\\[4.00mm]
%%\beqa
\hspace*{-2.000mm}
\left. \left\{\widehat{\eta}^{\,6}_{1}\,, \ldots \,,\, \widehat{\eta}^{\,6}_{15}\right\} \right|_\text{$\kappa$-realization}&=&
%%6
\Big\{-\frac{143}{250},\,\,\frac{2}{5},\,\frac{921}{1000},\,
-\frac{313}{500},\,-\frac{603}{1000},\,-\frac{449}{1000},\,
\frac{609}{1000},\,-\frac{39}{1000},\,
\nonumber\\[1mm] \hspace*{-2.000mm}&&
\frac{443}{1000},\,\frac{383}{500},\,-\frac{17}{500},\,
\frac{27}{1000},\,\frac{181}{500},\,\frac{941}{1000},\,
-\frac{181}{1000}\Big\}\,,
\eeqa
%%\\[4.00mm]
\beqa
\hspace*{-2.000mm}
\left. \left\{\widehat{\eta}^{\,7}_{1}\,, \ldots \,,\, \widehat{\eta}^{\,7}_{15}\right\} \right|_\text{$\kappa$-realization}&=&
%%7
\Big\{-\frac{129}{250},\,-\frac{147}{1000},\,\frac{387}{1000},\,
-\frac{611}{1000},\,\frac{313}{1000},\,\frac{191}{200},\,
-\frac{61}{250},\,-\frac{11}{25},\,
\nonumber\\[1mm] \hspace*{-2.000mm}&&
-\frac{969}{1000},\,\frac{927}{1000},\,\frac{489}{500},\,
\frac{361}{500},\,\frac{57}{1000},\,-\frac{4}{125},\,
\frac{83}{100}\Big\}\,,
%%\eeqa
\\[4.00mm]
%%\beqa
\hspace*{-2.000mm}
 \left. \left\{\widehat{\eta}^{\,8}_{1}\,, \ldots \,,\, \widehat{\eta}^{\,8}_{15}\right\} \right|_\text{$\kappa$-realization}&=&
%%8
\Big\{-\frac{273}{1000},\,-\frac{441}{1000},\,\frac{427}{500},\,
-\frac{317}{1000},\,-\frac{407}{1000},\,-\frac{57}{200},\,
\frac{23}{1000},\,\frac{93}{200},\,
\nonumber\\[1mm] \hspace*{-2.000mm}&&
\frac{163}{250},\,\frac{36}{125},\,\frac{333}{1000},\,
-\frac{18}{25},\,\frac{321}{1000},\,\frac{307}{500},\,
-\frac{193}{1000}\Big\}\,,
\eeqa
%%\\[4.00mm]
\beqa
\hspace*{-2.000mm}
\left. \left\{\widehat{\eta}^{\,9}_{1}\,, \ldots \,,\, \widehat{\eta}^{\,9}_{15}\right\} \right|_\text{$\kappa$-realization}&=&
%%9
\Big\{\frac{1}{250},\,-\frac{2}{125},\,-\frac{981}{1000},\,
\frac{112}{125},\,-\frac{137}{500},\,\frac{547}{1000},\,
-\frac{201}{500},\,\frac{101}{500},\,
\nonumber\\[1mm] \hspace*{-2.000mm}&&
-\frac{63}{250},\,\frac{62}{125},\,\frac{597}{1000},\,
-\frac{114}{125},\,-\frac{39}{200},\,\frac{197}{200},\,
-\frac{7}{100}\Big\}\,,
%%\eeqa
\\[4.00mm]
%%\beqa
\hspace*{-2.000mm}
\left. \left\{\widehat{\eta}^{\,10}_{1}\,, \ldots \,,\,
\widehat{\eta}^{\,10}_{15}\right\} \right|_\text{$\kappa$-realization}&=&
%%10
\Big\{\frac{969}{1000},\,\frac{4}{5},\,-\frac{243}{1000},\,
\frac{29}{250},\,-\frac{38}{125},\,-\frac{151}{500},\,
\frac{17}{25},\,\frac{119}{1000},\,
\nonumber\\[1mm] \hspace*{-2.000mm}&&
\frac{463}{1000},\,-\frac{33}{200},\,\frac{257}{500},\,
\frac{219}{250},\,-\frac{319}{1000},\,\frac{119}{500},\,
\frac{577}{1000}\Big\}\,.
\eeqa
\esubeqs
The corresponding matrices $\widehat{\eta}^{\,\mu}_\text{$\kappa$-realization}$
have been given in App.~B of Ref.~\cite{Klinkhamer2021-numsol}.

Remark that, following Ref.~\cite{Klinkhamer2021-first-look},
we have chosen \emph{rational} numbers for the random constants
in \eqref{eq:phat-kappa-realization}
and \eqref{eq:etahatcoeff-kappa-realization}.
The reason is that we can then easily write
down their \emph{exact} values, whereas arbitrary real numbers would require
an infinite number of digits (or implicit defining relations,
such as for the irrational numbers $\sqrt{3}$ and $\pi$\,).

%%\newpage%%tmp
\section{Coefficients from the full algebraic equation for (D, N) = (10, 4)}
\label{app:Coefficients-from-full-algebraic-equation}

Denoting the approximate numerical solution from the last row  of
Table~\ref{tab-num-sols-full} by ``$\kappa\text{-\underline{num-sol}}$'',
the 300 real numbers defining the matrices
\eqref{eq:ahat-matrices-kappa-num-sol}  are
[displaying ten batches of $30$ numbers each]:
\beqa\label{eq:300-coeffs-f-33pt8776}
&&
\left.
\left\{
\widehat{r}_{1}^{\;1}\,,\, \widehat{s}_{1}^{\;1}\,,
\,\ldots\,,\,
\widehat{r}_{1}^{\;15}\,,\, \widehat{s}_{1}^{\;15}\,,
\,
\widehat{r}_{2}^{\;1}\,,\, \widehat{s}_{2}^{\;1}\,,
\,\ldots\,,\,
\widehat{r}_{10}^{\;15}\,,\, \widehat{s}_{10}^{\;15}
\right\}\right|_{\kappa\text{-\underline{num-sol}}}^{(f_\text{penalty}=33.8776)}
=
\nonumber\\&&%%
\nonumber\\[1mm]&&
\big\{  %%\nonumber\\&&%%
-0.0692546,-0.58643,-1.05642,-0.601179,-0.148017,0.607459,   %%row1
\nonumber\\&&%%
-0.295285,-0.140475,-0.59204,0.274455,-0.508568,0.309931,
\nonumber\\&&%%
-0.0620355,0.381277,0.211426,0.178033,-0.112937,-0.138966,
\nonumber\\&&%%
0.469841,0.463514,-0.00762073,-0.155881,0.0143421,0.466782,
\nonumber\\&&%%
0.580538,0.439774,0.0995092,-0.622271,-0.68924,-0.437244,
\nonumber\\&&%%
\nonumber\\&&%%
0.426018,-0.509828,-0.439085,0.293402,0.486298,-0.658821,
\nonumber\\&&%%
-0.219118,0.0364208,0.824188,0.429625,0.276073,0.179746,
\nonumber\\&&%%
-0.068707,0.0925217,-0.118121,-0.186855,-0.00933079,0.219118,
\nonumber\\&&%%
0.520674,0.762843,0.0327783,0.240407,-0.730098,-0.499508,
\nonumber\\&&%%
-0.0487667,0.269917,0.328138,0.133357,-0.226639,-0.21064,   %%row10
\nonumber\\&&%%
\nonumber\\&&%%
0.266147,0.0693405,-0.0752673,0.324464,-0.468063,-0.537566,
\nonumber\\&&%%
-0.522858,-0.719695,-0.0162757,-0.289673,0.0600267,-0.473248,
\nonumber\\&&%%
0.565031,0.0352853,0.609935,0.341532,0.251534,-0.141831,
\nonumber\\&&%%
-0.145027,-0.146535,0.020147,0.0975357,0.143888,-0.389464,
\nonumber\\&&%%
0.00800391,-0.44477,-0.491185,-0.120212,-0.521562,0.0655693,
\nonumber%\\&&%%
\eeqa
\beqa
%\nonumber\\
&&%%
-0.769702,0.393627,0.0849714,-0.57002,-0.00425748,0.154496,
\nonumber\\&&%%
0.487595,-0.0372879,-0.342562,0.289481,-0.0750309,0.611541,
\nonumber\\&&%%
0.163813,-0.17035,-0.0421598,0.165941,0.451901,-0.353845,
\nonumber\\&&%%
-0.065826,0.708222,0.294372,0.425108,0.384654,0.0226756,
\nonumber\\&&%%
0.570911,-0.324845,-0.261706,0.035435,-0.483642,0.251261,  %%row20
\nonumber\\&&%%
\nonumber\\&&%%
-0.342193,-0.50229,0.125031,-0.64019,-0.466247,0.53052,
\nonumber\\&&%%
0.275081,-0.730265,0.853657,-0.121877,-0.85128,-0.246815,
\nonumber\\&&%%
0.0756539,0.120902,0.268202,0.403389,-0.685643,0.503552,
\nonumber\\&&%%
0.0864294,-0.224326,0.352859,-0.248242,-0.0473426,0.0953277,
\nonumber\\&&%%
0.23623,0.39878,0.407025,0.376143,0.0485194,0.0974565, %%row25
\nonumber\\&&%%
\nonumber\\&&%%
-0.241239,0.0523708,-0.0378564,0.394186,-0.0230116,0.285019,  %%row26
\nonumber\\&&%%
0.376322,-0.14455,0.141922,0.284971,0.214979,-0.650774,
\nonumber\\&&%%
0.160804,-0.0462938,0.325296,0.337869,-0.0518918,-0.273317,
\nonumber\\&&%%
-0.44725,-0.187702,-0.452864,-0.00378001,-0.388691,0.358931,
\nonumber\\&&%%
0.382993,-0.866705,0.255596,0.631997,0.013364,-0.311463,  %%row30
\nonumber\\&&%%
\nonumber\\&&%%
-0.457347,0.289158,-0.344167,0.0958883,0.246078,-0.254109,
\nonumber\\&&%%
-0.68349,0.111889,0.391893,-0.360248,-0.0215233,-0.259903,
\nonumber\\&&%%
0.543266,0.0653004,-0.653161,0.270118,-0.436127,0.19117,
\nonumber\\&&%%
-0.588626,0.188583,0.684876,-0.0178182,0.323925,-0.412551,
\nonumber\\&&%%
0.255511,-0.197965,-0.25607,0.110931,0.192163,-0.111291,
\nonumber\\&&%%
\nonumber\\&&%%
-0.141328,0.270626,-0.417135,0.0142941,-0.524982,-0.269947,
\nonumber\\&&%%
-0.636919,-0.237791,0.367274,0.809111,0.202189,0.314854,
\nonumber\\&&%%
-0.237622,0.229124,-0.472898,0.190654,-0.216502,0.559077,
\nonumber\\&&%%
0.18242,0.657434,0.719802,-0.307403,-0.0636525,-0.511735,
\nonumber\\&&%%
0.43477,0.231476,0.832258,0.131869,0.436263,0.666609,  %%row40
\nonumber\\&&%%
\nonumber\\&&%%
-0.1219,-0.0734223,0.29402,0.248577,-0.101651,-0.0870295,
\nonumber\\&&%%
-0.315557,0.292283,-0.347289,-0.0371514,-0.265792,0.290594,
\nonumber\\&&%%
0.140305,0.45099,0.296089,0.392803,-0.0969607,-0.0578319,
\nonumber\\&&%%
0.62338,0.361966,0.422547,0.570988,-0.588562,0.41253,
\nonumber\\&&%%
0.0313179,0.0259956,-0.361538,-0.730508,-0.43556,0.133992,
\nonumber %\\&&%%
\eeqa
\beqa
%\nonumber\\
&&%%
-0.494957,-0.389868,-0.953034,-0.177046,-0.358249,0.186803,
\nonumber\\&&%%
-0.522166,0.0464763,0.355931,-0.048642,-0.105042,-0.283861,
\nonumber\\&&%%
-0.216209,-0.0980749,0.622007,0.707796,0.204977,-0.505334,
\nonumber\\&&%%
-0.233755,0.7281,0.372945,0.588971,-0.176769,0.422182,
\nonumber\\&&%%
-0.102989,0.366496,-0.743647,0.526077,-0.350589,-0.045876  %%row50
\big\}\,,
\eeqa
where up to 6 significant digits are shown.
The superscript with the
$f_\text{penalty}$ value has been omitted in the main text.

Just in order to avoid any misunderstanding: the above 300 numbers
are only given for illustrative purposes, because they are, most likely,
different from the correct values (as mentioned in the last paragraph
of  Sec.~\ref{subsec:Num-results-full-alg-eq}).

%%\newpage%%tmp
%%\section{Nontrivial critical points for the case $\mathbf{(D,\,N)=(3,\,3)}$}
\section{Nontrivial critical points for the case (D, N) = (3, 3)}
\label{app:Nontrivial-critical-points}
%%eqs from sols-bos-master-eq-v5 final version

\subsection{Critical-point setup: General case}
\label{app:Critical-point-setup-General-case}

This appendix aims at being more or less self-contained.
Consider a generalized version of the IIB matrix
model~\cite{IKKT-1997,Aoki-etal-review-1999} with
a different number of bosonic matrices
($D=3,\,4,\,6,\,10$) and different matrix sizes ($N\geq 2$), whereas
the genuine model has $D=10$ and $N \gg 1$.
After the fermionic matrices have been integrated out,
the partition function reads~\cite{IKKT-1997,Aoki-etal-review-1999,KrauthNicolaiStaudacher1998}%
\bsubeqs\label{eq:app-Z-D-N-with-Seff-all-defs}
\beqa
\label{eq:app-Z-D-N-with-Seff}
\hspace*{-0.0mm}
Z_{D,\,N} &=&
\int \prod_{c=1}^{g}\,\prod_{\mu=1}^{D}\,
d A_{\mu}^{c}\;
%%\exp\left( -\,S_{\text{eff},\,D,\,N}[A] \right)\,,
e^{\displaystyle{-\,S_{\text{eff},\,D,\,N}[A]}}\,,
\\[2mm]
\label{eq:app-Seff-D-N}
\hspace*{-0.0mm}
S_{\text{eff},\,D,\,N}[A]
&=&
S_{\text{bos},\,D,\,N}[A]- \log\,\mathcal{P}_{D,\,N}[A]\,,
\\[2mm]
\hspace*{-0.0mm}
\label{eq:app-Sbos}
S_{\text{bos},\,D,\,N}[A] &=&
-\frac{1}{2}\,\text{Tr}\,
\Big(\big[A^{\mu},\,A^{\nu} \big]\,\big[ A^{\mu},\,A^{\nu} \big]\,\Big)\,,
%%\eeqa
\\[2mm]
%%\beqa
\hspace*{-0.0mm}
\label{eq:app-Amu-coeff}
A_{\mu}
&=&A_{\mu}^{c}\,t_{c}\,,
\quad
A_{\mu}^{c} \in \mathbb{R}\,,
\quad
t_{c} \in \text{su}(N)\,,
\\[2mm]
\hspace*{-0.0mm}
\label{eq:app-trace-T-T}
\text{Tr}\, \big( t_{c} \cdot t_{d} \big)
&=& \frac{1}{2}\;\delta_{c d}\,,
%%\eeqa
\\[2mm]
%%\beqa
\label{eq:app-g}
\hspace*{-0.0mm}
g &\equiv& N^2-1\,,
\eeqa
\esubeqs
where repeated Greek indices are summed over
(just as having an implicit Euclidean ``metric'')
and the quantity $\mathcal{P}_{D,\,N}$ in \eqref{eq:app-Seff-D-N}
will be discussed shortly.
The commutators entering the bosonic action term \eqref{eq:app-Sbos}
are defined by $[X,\,Y]\equiv$ $X \cdot Y - Y \cdot X$
for square ma\-tri\-ces $X$ and $Y$ of equal dimension.
The expansion \eqref{eq:app-Amu-coeff},
for real coefficients $A_{\mu}^{c}$,
uses the $N \times N$ traceless Hermitian
$SU(N)$ generators $t_{c}$ with normalization \eqref{eq:app-trace-T-T}.

The Gaussian-type integration of the fermionic matrices,
for $D = 3,\,4,\,10$,
produces the Pfaffian $\mathcal{P}_{D,\,N}[A]$,
which is given explicitly by a sum over permutations
or by a sum involving the Levi--Civita symbol.
Including the $D = 6$ case, the quantity $\mathcal{P}_{D,\,N}[A]$
is a homogenous polynomial in the bosonic coefficients
$A_{\mu}^{c}$, where the order $K$,
for values $D\in \{3,\,4,\,6,\,10\}$ with supersymmetry,
is given by
\beq
\label{eq:app-Pfaffian-order-polynomial-K}
K =
\big( D-2\big)\,\big( N^2-1\big)\,.
\eeq
Further discussion of the polynomial $\mathcal{P}_{D,\,N}[A]$ appears in
Sec.~\ref{subsec:Partition-function}
and Refs.~\cite{KrauthNicolaiStaudacher1998,NishimuraVernizzi2000-JHEP}.
%the research papers~\cite{KrauthNicolaiStaudacher1998,NishimuraVernizzi2000-JHEP}
%and a brief summary has also been given in App.~A of Ref.~\cite{Klinkhamer2021-numsol}.

The issue of the convergence of
the integrals in \eqref{eq:app-Z-D-N-with-Seff}
has been studied by the authors of
Ref.~\cite{AustingWheater2001},
with the conclusion that there is
absolute convergence for $D=4,\, 6,\, 10$.
In any case, it may be of mathematical interest
to look for the critical points (even for $D=3$)
and, more precisely, to establish their existence.
Incidentally, critical points of the matrix model
have also been discussed in a recent paper~\cite{Steinacker2022},
which, however, appears to consider only
critical points of the bosonic action \eqref{eq:app-Sbos}.

Here, we present results for the existence of nontrivial critical points
of the effective bosonic action \eqref{eq:app-Seff-D-N}.
Specifically, we get explicit solutions
for a special case with low values of $D$ and $N$.
Remark that this effective action incorporates
the fermionic ``quantum fluctuations''  exactly.

%%\newpage%%tmp
%%\subsection{Critical-point setup: Special case with $(D,\,N)=(3,\,3)$}
\subsection{Critical-point setup: Special case with (D, N) = (3, 3)}
\label{app:Critical-point-setup-Special-case}

Consider the matrix model \eqref{eq:app-Z-D-N-with-Seff-all-defs}
with the following parameters:
\beq
\label{eq:app-D3-N3}
\big\{ D,\,  N \big\}  = \big\{ 3,\,  3 \big\}\,.
\eeq
The eight generators $t_{c}$ are proportional to the
$3 \times 3$ Gell-Mann matrices $\lambda_{c}$
from the ``eightfold-way'' (1961) and we take explicitly
%\bsubeqs
%\beqa
\begin{align}
\label{eq:app-SU3-generators}
\widehat{t}_{1} &=  \frac{1}{2}\,
\left(
\begin{array}{ccc}
 0 & 1 & 0 \\
 1 & 0 & 0 \\
 0 & 0 & 0 \\
\end{array}
\right)\,,
\quad
&\widehat{t}_{2} &= \frac{1}{2}\,
\left(
\begin{array}{ccc}
 0 & i & 0 \\
 -i & 0 & 0 \\
 0 & 0 & 0 \\
\end{array}
\right)\,,
\quad
&\widehat{t}_{3} &= \frac{1}{2}\,
\left(
\begin{array}{ccc}
 0 & 0 & 1 \\
 0 & 0 & 0 \\
 1 & 0 & 0 \\
\end{array}
\right)\,,
\nonumber\\[2mm]
\widehat{t}_{4} &= \frac{1}{2}\,
\left(
\begin{array}{ccc}
 0 & 0 & i \\
 0 & 0 & 0 \\
 -i & 0 & 0 \\
\end{array}
\right)\,,
\quad
&\widehat{t}_{5} &=\frac{1}{2}\,
\left(
\begin{array}{ccc}
 0 & 0 & 0 \\
 0 & 0 & 1 \\
 0 & 1 & 0 \\
\end{array}
\right)\,,
\quad
&\widehat{t}_{6} &=\frac{1}{2}\,
\left(
\begin{array}{ccc}
 0 & 0 & 0 \\
 0 & 0 & i \\
 0 & -i & 0 \\
\end{array}
\right)\,,
\nonumber\\[2mm]
\widehat{t}_{7} &=\frac{1}{2}\,
\left(
\begin{array}{ccc}
 1 & 0 & 0 \\
 0 & -1 & 0 \\
 0 & 0 & 0 \\
\end{array}
\right)\,,
\quad
&\widehat{t}_{8} &=\frac{1}{2 \sqrt{3}}\,
\left(
\begin{array}{ccc}
 1 & 0 & 0 \\
 0 & 1 & 0 \\
 0 & 0 & -2 \\
\end{array}
\right)\,, & &
\end{align}
%\eeqa
%\esubeqs
%\begin{align}
%x&=y           &  w &=z              &  a&=b+c \nonumber\\
%2x&=-y         &  3w&=\frac{1}{2}z   &  a&=b   \nonumber\\
%-4 + 5x&=2+y   &  w+2&=-1+w          &  &
%\end{align}
where the hat distinguishes these generators from those in
App.~\ref{app:SU4-generators}.

The main reason for considering this special case is that
there is now an explicit compact result
for the Pfaffian~\cite{KrauthNicolaiStaudacher1998}:
\beqa
\label{eq:app-Pfaffian-D3-N3}
\hspace*{-10mm}
\mathcal{P}_{3,\,3}[A] &=&
-\frac{3}{4}\,
\text{Tr}\,\Big(
\big[A^{\mu},\,A^{\nu} \big]\,\big\{ A^{\rho},\,A^{\sigma} \big\}\,
\Big)\;
\text{Tr}\,\Big(
\big[A^{\mu},\,A^{\nu} \big]\,\big\{ A^{\rho},\,A^{\sigma} \big\}\,
\Big)
\nonumber\\[1mm]
\hspace*{-10mm}
&& +
\frac{6}{5}\,\text{Tr}\,
\Big(
A^{\mu}\,\big[ A^{\nu},\,A^{\rho} \big]\,\Big)\
\text{Tr}\,\Big(
A^{\mu}\,\big[ \big\{ A^{\nu},\,A^{\sigma} \big\},\,
              \big\{ A^{\rho},\,A^{\sigma} \big\}\big]\,\Big)\,,
\eeqa
which corresponds to a homogenous eighth-order polynomial
in the bosonic coefficients $A_{\mu}^{c}$.
This expression also contains anticommutators,
defined by $\{X,\,Y\}\equiv$ $X \cdot Y + Y \cdot X$
for square ma\-tri\-ces $X$ and $Y$ of equal dimension.

The critical points are then obtained from the
following algebraic equation:
\beqa
\label{eq:app-algebraic-equation-D3-N3}
\frac{\delta S_{\text{eff},\,3,\,3}\big[\,A\,\big]}
      {\delta\, A^{\mu}_{lk}}
&=&
\Big[ A^{\,\nu},\,\big[A^{\,\nu},\,A^{\;\mu}\big]\Big]_{kl}
- \frac{1}{\mathcal{P}_{3,\,3}\left(A\right)}\;
\frac{\partial\, \mathcal{P}_{3,\,3}\left(A\right)}
     {\partial\, A^{\;\mu}_{\;lk}}
= 0             \,,
\eeqa
with expression \eqref{eq:app-Pfaffian-D3-N3} for the Pfaffian.
Multiplying \eqref{eq:app-algebraic-equation-D3-N3}
by $\mathcal{P}_{3,\,3}\left(A\right)$
gives an $11$-th order homogeneous polynomial.

Explicit solutions of \eqref{eq:app-algebraic-equation-D3-N3}
can be obtained from the procedure of Ref.~\cite{Klinkhamer2021-sols-D3-N3}
(see, in particular, the last paragraph of Sec.~4.1 in that reference)
by setting the master momenta and the master noise there to zero,
$\widehat{p}_{k}=0$ and $\widehat{\eta}^{\;\mu}_{\;kl}=0$.
Very briefly, the procedure is to build a penalty function as the
sum of the squares of the 24 component equations
(without further overall numerical factors)
and to use the numerical minimization routine \texttt{FindMinimum}
from \textsc{Mathematica} 12.1 (cf. Ref.~\cite{Wolfram1991}).
Typically, we use a 36-digit working precision
(see Sec.~\ref{app:Technical-remarks}
for further comments).

%%\newpage%%tmp
%%\subsection{Critical-point results for $(D,\,N)=(3,\,3)$}
\subsection{Critical-point results for (D, N) = (3, 3)}
\label{app:Critical-point-results}

\subsubsection{First critical-point solution}
\label{app:First-critical-point-solution}

The coefficients of a first nontrivial
solution (denoted by an overbar)
of the algebraic equation \eqref{eq:app-algebraic-equation-D3-N3}
are:
\bsubeqs\label{eq:app-sol1-coeff}
\beqa
\hspace*{-8mm}
\left\{\!
\begin{array}{cc}
\overline{A}_{1}^{\,1}\,, & \overline{A}_{1}^{\,2}\\
\overline{A}_{1}^{\,3}\,, & \overline{A}_{1}^{\,4}\\
\overline{A}_{1}^{\,5}\,, & \overline{A}_{1}^{\,6}\\
\overline{A}_{1}^{\,7}\,, & \overline{A}_{1}^{\,8}\\
\end{array}
\!\right\}
&\!\!=\!\!&
\left\{\!
\renewcommand{\arraycolsep}{0.pc} %% enlarge column spacing
\renewcommand{\arraystretch}{1.25}  %% enlarge line spacing
\begin{array}{cc}
0.570574083490128575476108\,,  &   0.658442593948256740170834\\
0.591172615092998049901068\,,  &   0.459965150810118963329943\\
0.0276034295370299604330960\,, & -0.662039140319935188116796\\
-0.151421059211537951401624\,, &  1.03331137617662167921442\\
\end{array}
\right\}\!,
\\[4mm]
\hspace*{-8mm}
\left\{\!
\begin{array}{cc}
\overline{A}_{2}^{\,1}\,, & \overline{A}_{2}^{\,2}\\
\overline{A}_{2}^{\,3}\,, & \overline{A}_{2}^{\,4}\\
\overline{A}_{2}^{\,5}\,, & \overline{A}_{2}^{\,6}\\
\overline{A}_{2}^{\,7}\,, & \overline{A}_{2}^{\,8}\\
\end{array}
\!\right\}
&\!\!=\!\!&
\left\{\!
\renewcommand{\arraycolsep}{0.pc} %% enlarge column spacing
\renewcommand{\arraystretch}{1.25}  %% enlarge line spacing
\begin{array}{cc}
0.809211845911449987314691\,, & 0.833691639722612054210554 \\
0.617000774572453785555974\,, & -0.164275893241917159317804 \\
0.830645113307734538465843\,, & -0.340131664055950118091891 \\
0.644115685767825480672465\,, & 1.54439672938962116994917 \\
\end{array}
\right\}\!,
\\[4mm]
\hspace*{-8mm}
\left\{\!
\begin{array}{cc}
\overline{A}_{3}^{\,1}\,, & \overline{A}_{3}^{\,2}\\
\overline{A}_{3}^{\,3}\,, & \overline{A}_{3}^{\,4}\\
\overline{A}_{3}^{\,5}\,, & \overline{A}_{3}^{\,6}\\
\overline{A}_{3}^{\,7}\,, & \overline{A}_{3}^{\,8}\\
\end{array}
\!\right\}
&\!\!=\!\!&
\left\{\!
\renewcommand{\arraycolsep}{0.pc} %% enlarge column spacing
\renewcommand{\arraystretch}{1.25}  %% enlarge line spacing
\begin{array}{cc}
0.504330365038665104746314\,,  & 1.18338120963105512893747  \\
0.0879652655029969016881395\,, & 0.416121944105811292070174 \\
0.614213738185278850877443\,,  & -0.945948255419088870571396 \\
0.811508170531654323184674\,,  & 0.926928461994400651864667   \\
\end{array}
\right\}\!,
\eeqa
\esubeqs
where only 24 significant digits have been shown.
All information of this particular solution is contained
in the 24 real numbers $\overline{A}_{\mu}^{\,c}$.
Let us now have a closer look at what the nature of this solution is,
while showing less digits than above.

From \eqref{eq:app-sol1-coeff}, we have the following three matrices:
\bsubeqs\label{eq:app-sol1-matrices}
\beqa
\overline{A}_{1}
&=&
\left(
\renewcommand{\arraycolsep}{0.5pc} %% enlarge column spacing
\renewcommand{\arraystretch}{1.5}  %% enlarge line spacing
\begin{array}{ccc}
 0.223 & 0.285+0.329\, i & 0.296+0.230\, i \\
 0.285-0.329\, i & 0.374 & 0.014-0.331\, i \\
 0.296-0.230\, i & 0.014+0.331\, i & -0.597 \\
\end{array}
\right)\,,\\[2mm]
\overline{A}_{2}
&=&
\left(
\renewcommand{\arraycolsep}{0.5pc} %% enlarge column spacing
\renewcommand{\arraystretch}{1.5}  %% enlarge line spacing
\begin{array}{ccc}
 0.768 & 0.405+0.417\, i & 0.309-0.082\, i \\
 0.405-0.417\, i & 0.124 & 0.415-0.170\, i \\
 0.309+0.082\, i & 0.415+0.170\, i & -0.892 \\
\end{array}
\right)\,,
%%\eeqa
\\[2mm]
%%\beqa
\overline{A}_{3}
&=&
\left(
\renewcommand{\arraycolsep}{0.5pc} %% enlarge column spacing
\renewcommand{\arraystretch}{1.5}  %% enlarge line spacing
\begin{array}{ccc}
 0.673 & 0.252+0.592\, i & 0.044+0.208\, i \\
 0.252-0.592\, i & -0.138 & 0.307-0.473\, i \\
 0.044-0.208\, i & 0.307+0.473\, i & -0.535 \\
\end{array}
\right)\,,
\eeqa
\esubeqs
and the corresponding matrices with absolute values of
the entries (in a short-hand notation):
\bsubeqs\label{eq:app-sol1-matrices-absolute-values}
\beqa
\text{Abs}\left[\,\overline{A}_{1}\right]
&=&
\left(
\renewcommand{\arraycolsep}{0.5pc} %% enlarge column spacing
\renewcommand{\arraystretch}{1.5}  %% enlarge line spacing
\begin{array}{ccc}
 0.223 & 0.436 & 0.375 \\
 0.436 & 0.374 & 0.331 \\
 0.375 & 0.331 & 0.597 \\
\end{array}
\right)\,,
\\
[2mm]
\text{Abs}\left[\,\overline{A}_{2}\right]
&=&
\left(
\renewcommand{\arraycolsep}{0.5pc} %% enlarge column spacing
\renewcommand{\arraystretch}{1.5}  %% enlarge line spacing
\begin{array}{ccc}
 0.768 & 0.581 & 0.319 \\
 0.581 & 0.124 & 0.449 \\
 0.319 & 0.449 & 0.892 \\
\end{array}
\right)\,,
\\[2mm]
\text{Abs}\left[\,\overline{A}_{3}\right]
&=&
\left(
\renewcommand{\arraycolsep}{0.5pc} %% enlarge column spacing
\renewcommand{\arraystretch}{1.5}  %% enlarge line spacing
\begin{array}{ccc}
 0.673 & 0.643 & 0.213 \\
 0.643 & 0.138 & 0.564 \\
 0.213 & 0.564 & 0.535 \\
\end{array}
\right)\,.
\eeqa
\esubeqs
Inspection of the matrices \eqref{eq:app-sol1-matrices-absolute-values},
in particular, shows that the far-off-diagonal
entries at positions $[1,\,3]$ and $[3,\,1]$
are not especially small.

We can quantify this conclusion by
calculating the average band-diagonal value from 2+3+2 entries
in \eqref{eq:app-sol1-matrices-absolute-values}
and the average off-band-diagonal value from 1+1 entries.
Then, determine the ratio $\widehat{R}_{\mu}$ of the average band-diagonal value
over the average off-band-diagonal value.
Summarizing this procedure by an equation, we have
the following definition of the ratio $\widehat{R}$ for a symmetric
$3\times 3$ matrix $\widehat{M}$
with nonnegative entries $\widehat{m}[i,\,j]$:
\beq
\label{eq:app-def-Rhat}
\widehat{R} \equiv
\frac{1}{7}\,\left( \sum_{i=1}^{3} \widehat{m}[i,\,i]
+2\,\sum_{j=1}^{2} \widehat{m}[j,\,j+1]\right)\;
\frac{1}{\widehat{m}[1,\,3]}\,,
\eeq
where we have used the symmetry of $M$ to simplify the expression.
In this way, we get,
from the three matrices \eqref{eq:app-sol1-matrices-absolute-values},
the following three ratios:
\beqa\label{eq:app-Abar-ratios}
\hspace*{-0mm}
\Big\{
\widehat{R}_{1},\,\widehat{R}_{2},\,\widehat{R}_{3}
\Big\}_{\text{Abs}\left[\,\overline{A}_{\mu}\right]}
&\!=\!&
\left\{ 1.04,\,  1.72,\,  2.53 \right\}\,,
\eeqa
which are all of order unity.

Following earlier work~\cite{Klinkhamer2021-first-look},
we diagonalize and order $\overline{A}_{1}$
(the transformed matrices are denoted by a prime) and get
\bsubeqs\label{eq:app-sol1-matrices-prime}
\beqa
\overline{A}_{1}^{\,\prime}
&=&
\left(
\renewcommand{\arraycolsep}{0.5pc} %% enlarge column spacing
\renewcommand{\arraystretch}{1.5}  %% enlarge line spacing
\begin{array}{ccc}
 -0.835 & 0 & 0 \\
 0 & -0.0169 & 0 \\
 0 & 0 & 0.852 \\
\end{array}
\right)\,,\\[2mm]
\overline{A}_{2}^{\,\prime}
&=&
\left(
\renewcommand{\arraycolsep}{0.5pc} %% enlarge column spacing
\renewcommand{\arraystretch}{1.5}  %% enlarge line spacing
\begin{array}{ccc}
 -0.719 & -0.097+0.481\, i & -0.0586+0.0059\, i \\
 -0.097-0.481\, i & -0.356 & 0.362+0.256\, i \\
 -0.0586-0.0059\, i & 0.362-0.256\, i & 1.08 \\
\end{array}
\right)\,,\\[2mm]
\overline{A}_{3}^{\,\prime}
&=&
\left(
\renewcommand{\arraycolsep}{0.5pc} %% enlarge column spacing
\renewcommand{\arraystretch}{1.5}  %% enlarge line spacing
\begin{array}{ccc}
 -0.658 & -0.466+0.088\, i & -0.0484-0.0255\, i \\
 -0.466-0.088\, i & -0.326 & 0.060+0.413\, i \\
 -0.0484+0.0255\, i & 0.060-0.413\, i & 0.984 \\
\end{array}
\right)\,.
\eeqa
\esubeqs
It is now clear that, for $\overline{A}_{2}^{\,\prime}$
and $\overline{A}_{3}^{\,\prime}$,  the far-off-diagonal
entries at positions $[1,\,3]$ and $[3,\,1]$
have rather small absolute values.
The corresponding ratios are:
\beqa\label{eq:app-Abar-prime-ratios}
\hspace*{-0mm}
\Big\{
\widehat{R}_{1},\,\widehat{R}_{2},\,\widehat{R}_{3}
\Big\}_{\text{Abs}\left[\,\overline{A}_{\mu}^{\,\prime}\right]}
&\!=\!&
\left\{ \infty,\, 9.76,\, 9.81\right\}\,,
\eeqa
where the last two ratios are of order ten.

Similarly, we can diagonalize and order $\overline{A}_{2}$
(the transformed matrices are denoted by a double prime) and we get
the following ratios:
\beqa\label{eq:app-Abar-primeprime-ratios}
\hspace*{-0mm}
\Big\{
\widehat{R}_{1},\,\widehat{R}_{2},\,\widehat{R}_{3}
\Big\}_{\text{Abs}\left[\,\overline{A}_{\mu}^{\,\prime\prime}\right]}
&\!=\!&
\left\{ 12.2,\, \infty,\, 13.1 \right\}\,,
\eeqa
where both nontrivial ratios are again of order ten.
The same result is also obtained
if we diagonalize and order $\overline{A}_{3}$
(the transformed matrices are denoted by a triple prime), with
the following ratios:
\beqa\label{eq:app-Abar-primeprimeprime-ratios}
\hspace*{-0mm}
\Big\{
\widehat{R}_{1},\,\widehat{R}_{2},\,\widehat{R}_{3}
\Big\}_{\text{Abs}\left[\,\overline{A}_{\mu}^{\,\prime\prime\prime}\right]}
&\!=\!&
\left\{ 11.3,\, 12.1,\, \infty\right\}\,.
\eeqa

The conclusion is that the matrix solution of
\eqref{eq:app-algebraic-equation-D3-N3}
with coefficients \eqref{eq:app-sol1-coeff}
has, upon diagonalization and ordering of one matrix,
a clear diagonal/band-diagonal structure,
even for the small matrix size considered, $N=3$.

%%\newpage%%tmp
\subsubsection{Second critical-point solution}
\label{app:Second-critical-point-solution}

\begin{figure}[t]
\begin{center}
\hspace*{0mm}
\includegraphics[width=0.55\textwidth]
{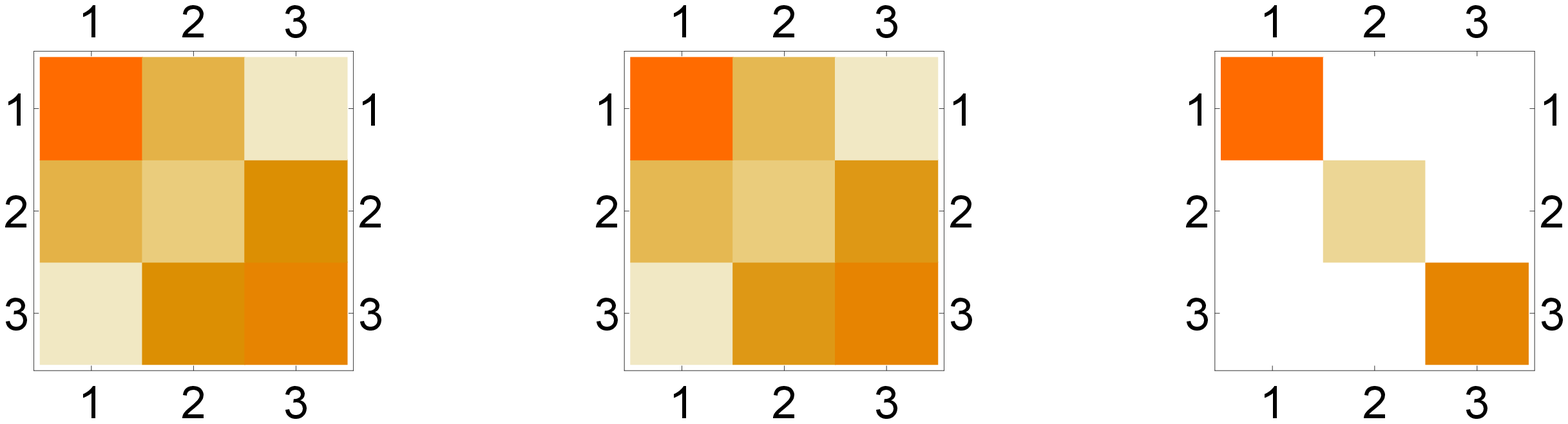}
%%{plot-critical-point-D3-N3-01mar2022new.eps}
\end{center}\vspace*{-0mm}
\caption{Density plots of
$\text{Abs}\big[\widetilde{A}_{\mu}^{\,\prime\prime\prime}\big]$
from the matrices \eqref{eq:app-sol2-matrices-primeprimeprime}.
Shown are $\mu=1,\,2\,,3$ from left to right.
}
\label{fig:Abs-Atilde-mu-primeprimeprime}
\vspace*{0mm}
\end{figure}

The coefficients of a second nontrivial
solution (denoted by a tilde)
of the algebraic equation \eqref{eq:app-algebraic-equation-D3-N3}
are:
\bsubeqs\label{eq:app-sol2-coeff}
\beqa
\hspace*{-12mm}
\left\{\!
\renewcommand{\arraycolsep}{0.pc} %% enlarge column spacing
\renewcommand{\arraystretch}{1.25}  %% enlarge line spacing
\begin{array}{cc}
\widetilde{A}_{1}^{\,1}\,,\, &\, \widetilde{A}_{1}^{\,2}\\
\widetilde{A}_{1}^{\,3}\,,\, &\, \widetilde{A}_{1}^{\,4}\\
\widetilde{A}_{1}^{\,5}\,,\, &\, \widetilde{A}_{1}^{\,6}\\
\widetilde{A}_{1}^{7}\,\,,\, &\, \widetilde{A}_{1}^{\,8}\\
\end{array}
\!\right\}
&\!\!=\!\!&
\left\{\!
\renewcommand{\arraycolsep}{0.pc} %% enlarge column spacing
\renewcommand{\arraystretch}{1.25}  %% enlarge line spacing
\begin{array}{cc}
-0.0486118439919790675351903\,,  &  0.786258758852912152288640 \\
0.0529067158425608509722208\,,   &  0.405878380158241357550552 \\
0.0894325197805553299305947\,,   &  -0.958079473974395296656234 \\
0.0123068275644060562414431\,,   & 0.590582765278256159281746 \\
\end{array}
\right\}\!,
\\[4mm]
\hspace*{-12mm}
\left\{\!
\renewcommand{\arraycolsep}{0.pc} %% enlarge column spacing
\renewcommand{\arraystretch}{1.25}  %% enlarge line spacing
\begin{array}{cc}
\widetilde{A}_{2}^{\,1}\,,\, &\, \widetilde{A}_{2}^{\,2}\\
\widetilde{A}_{2}^{\,3}\,,\, &\, \widetilde{A}_{2}^{\,4}\\
\widetilde{A}_{2}^{\,5}\,,\, &\, \widetilde{A}_{2}^{\,6}\\
\widetilde{A}_{2}^{\,7}\,,\, &\, \widetilde{A}_{2}^{\,8}\\
\end{array}
\!\right\}
&\!\!=\!\!&
\left\{\!
\renewcommand{\arraycolsep}{0.pc} %% enlarge column spacing
\renewcommand{\arraystretch}{1.25}  %% enlarge line spacing
\begin{array}{cc}
0.529762679031679212023180\,,  &  0.769580268233783935440839 \\
-0.233354688796790875044128\,, &  -0.541601511215317101360294 \\
0.331576461482468810651328\,,  &  -0.779354061134799362179733 \\
0.531179126831174470587476\,,  &  0.652651364568218169075602 \\
\end{array}
\right\}\!,
\\[4mm]
\hspace*{-12mm}
\left\{\!
\renewcommand{\arraycolsep}{0.pc} %% enlarge column spacing
\renewcommand{\arraystretch}{1.25}  %% enlarge line spacing
\begin{array}{cc}
\widetilde{A}_{3}^{\,1}\,,\, &\, \widetilde{A}_{3}^{\,2}\\
\widetilde{A}_{3}^{\,3}\,,\, &\, \widetilde{A}_{3}^{\,4}\\
\widetilde{A}_{3}^{\,5}\,,\, &\, \widetilde{A}_{3}^{\,6}\\
\widetilde{A}_{3}^{\,7}\,,\, &\, \widetilde{A}_{3}^{\,8}\\
\end{array}
\!\right\}
&\!\!=\!\!&
\left\{\!
\renewcommand{\arraycolsep}{0.pc} %% enlarge column spacing
\renewcommand{\arraystretch}{1.25}  %% enlarge line spacing
\begin{array}{cc}
0.0360931848378888826301440\,, &  1.62034088330609352509732 \\
-1.02598168958966357957916\,,  &  0.286848079464457062317327 \\
0.379578994127041469609087\,,  &  -1.37050518862587829073235 \\
1.42746943415983021599759\,, &  0.554733364863768377772749 \\
\end{array}
\right\}\!,
\eeqa
\esubeqs
where, again, only 24 significant digits have been  shown.
This second solution leads to the same conclusions as the
first solution. Very briefly, the ratios \eqref{eq:app-def-Rhat}
from the original matrices are:
\beqa\label{eq:app-Atilde-ratios}
\hspace*{-0mm}
\Big\{
\widehat{R}_{1},\,\widehat{R}_{2},\,\widehat{R}_{3}
\Big\}_{\text{Abs}\left[\,\widetilde{A}_{\mu}\right]}
&\!=\!&
\left\{ 1.70,\, 1.30,\, 1.28 \right\}\,,
\eeqa
and those of the transformed  %%diagonalized
matrices:
\bsubeqs\label{eq:app-Atilde-diagonalized-ratios}
\beqa
\label{eq:app-Atilde-prime-ratios}
\hspace*{-0mm}
\Big\{\widehat{R}_{1},\,\widehat{R}_{2},\,\widehat{R}_{3}
\Big\}_{\text{Abs}\left[\,\widetilde{A}_{\mu}^{\,\prime}\right]}
&\!=\!&
\left\{ \infty,\, 10.4,\, 9.04 \right\}\,,
\\[4mm]
\label{eq:app-Atilde-primeprime-ratios}
\hspace*{-0mm}
\Big\{\widehat{R}_{1},\,\widehat{R}_{2},\,\widehat{R}_{3}
\Big\}_{\text{Abs}\left[\,\widetilde{A}_{\mu}^{\,\prime\prime}\right]}
&\!=\!&
\left\{ 10.2,\, \infty ,\, 9.41 \right\}\,,
\\[4mm]
\label{eq:app-Atilde-primeprimeprime-ratios}
\hspace*{-0mm}
\Big\{\widehat{R}_{1},\,\widehat{R}_{2},\,\widehat{R}_{3}
\Big\}_{\text{Abs}\left[\,\widetilde{A}_{\mu}^{\,\prime\prime\prime}\right]}
&\!=\!&
\left\{ 16.8,\, 18.0,\, \infty \right\}\,.
\eeqa
\esubeqs
Purely for illustrative purposes, we give the matrices
with the strongest band-diagonal structure:
\bsubeqs\label{eq:app-sol2-matrices-primeprimeprime}
\beqa
\widetilde{A}_{1}^{\,\prime\prime\prime}
&=&
\left(
\renewcommand{\arraycolsep}{0.5pc} %% enlarge column spacing
\renewcommand{\arraystretch}{1.5}  %% enlarge line spacing
\begin{array}{ccc}
 -0.603 & -0.083-0.250\, i & -0.0029+0.0213\, i \\
 -0.083+0.250\, i & 0.178 & 0.369-0.147\, i \\
 -0.0029-0.0213\, i & 0.369+0.147\, i & 0.425 \\
\end{array}
\right)\,,\\[2mm]
\widetilde{A}_{2}^{\,\prime\prime\prime}
&=&
\left(
\renewcommand{\arraycolsep}{0.5pc} %% enlarge column spacing
\renewcommand{\arraystretch}{1.5}  %% enlarge line spacing
\begin{array}{ccc}
 -0.730 & 0.239-0.130\, i & -0.0223+0.0013\, i \\
 0.239+0.130\, i & 0.215 & 0.189+0.359\, i \\
 -0.0223-0.0013\, i & 0.189-0.359\, i & 0.515 \\
\end{array}
\right)\,,\\[2mm]
\widetilde{A}_{3}^{\,\prime\prime\prime}
&=&
\left(
\renewcommand{\arraycolsep}{0.5pc} %% enlarge column spacing
\renewcommand{\arraystretch}{1.5}  %% enlarge line spacing
\begin{array}{ccc}
 -1.53 & 0 & 0 \\
 0 & 0.228 & 0 \\
 0 & 0 & 1.30 \\
\end{array}
\right)\,,
\eeqa
\esubeqs
where the off-band-diagonal entries
in $\widetilde{A}_{1}^{\,\prime\prime\prime}$
and $\widetilde{A}_{2}^{\,\prime\prime\prime}$
are indeed rather small
(cf. Fig.~\ref{fig:Abs-Atilde-mu-primeprimeprime}).

%%\newpage%%tmp
\subsubsection{Technical remarks}
\label{app:Technical-remarks}  %%{app: Accuracy-and-precision}

The results presented here have been obtained from the
numerical minimization routine \texttt{FindMinimum}
of \textsc{Mathematica} 12.1 (cf. Ref.~\cite{Wolfram1991})
with a 36-digit working precision, but, for comparison, we have also
used a 24-digit working precision
and a 48-digit working precision. The maximum absolute value of the
equation residues drops with increasing working precision,
as well as the value of the penalty function $f_\text{penalty}$;
see Table~\ref{tab-sol1} for details.
Most importantly, the 24-digit numbers in \eqref{eq:app-sol1-coeff}
are reproduced by the 48-digit-working-precision calculation.

Following Ref.~\cite{Klinkhamer2021-sols-D3-N3},
we will call the solutions \eqref{eq:app-sol1-coeff}
and \eqref{eq:app-sol2-coeff} ``quasi-exact,''
as their number of digits can, in principle, be increased arbitrarily.

\begin{table}[t]
\vspace*{-0mm}
\begin{center}
\caption{First solution of the $(D,\,N)=(3,\,3)$ extremal
equation \eqref{eq:app-algebraic-equation-D3-N3}
as presented in Sec.~\ref{app:First-critical-point-solution}.
The residues of the 24 component equations
$\text{eq-}\widehat{A}_{\mu}^{\;c}$
are computed (they all vanish for a perfect solution).
The quantity $\text{MaxAbsRes}$ is the maximum of the absolute values
of these residues
and the function $f_\text{penalty}$ is the sum of their squared
absolute values.
The expression $\text{eq-}A_{\mu}^{\;c}$ follows
from the left-hand side of \eqref{eq:app-algebraic-equation-D3-N3}
by performing a matrix multiplication with $\widehat{t}_{c}$,
taking the trace, and multiplying the result by two
[here, the $\widehat{t}_{c}$ are the $SU(3)$ generators
from App.~\ref{app:Critical-point-setup-Special-case}].
The calculations use the \texttt{FindMinimum} routine
from \textsc{Mathematica} 12.1~\cite{Wolfram1991}
with working precision (WP), accuracy goal (AG),
and precision goal (PG)  as shown.
}
\vspace*{0.5\baselineskip}%%\vspace*{1\baselineskip}
\label{tab-sol1}
\renewcommand{\tabcolsep}{1.5pc}    %% enlarge column spacing
\renewcommand{\arraystretch}{1.75}   %% enlarge line spacing
\begin{tabular}{c||c|c}
\hline\hline
\{WP, AG, PG\} &
$f_\text{penalty}\equiv \sum \big|\text{eq-}A_{\mu}^{\;c}\big|^{2}$  &
$\text{MaxAbsRes} \equiv
\text{max} \big\{\big|\text{eq-}A_{\mu}^{\;c}\big| \big\}$  \\
\hline\hline
%$\{ 14,\,  7,\,  7\}$
%& $\text{O}\left( 10^{} \right)$   &  $\text{O}\left( 10^{} \right)$
%\\
$\{ 24,\,  12,\,  12\}$
 & $\text{O}\left( 10^{-28} \right)$   &  $\text{O}\left( 10^{-14} \right)$
 \\
$\{ 36,\,  24,\,  24\}$
& $\text{O}\left( 10^{-55} \right)$   &  $\text{O}\left( 10^{-28} \right)$
\\
$\{ 48,\,  36,\,  36\}$
& $\text{O}\left( 10^{-91} \right)$   &  $\text{O}\left( 10^{-46} \right)$  \\
\hline\hline
\end{tabular}
\end{center}
\vspace*{-0mm}
\end{table}

%%\newpage%%tmp
\subsection{Discussion of critical-point solutions}
\label{app:Discussion-critical-point-solutions}

The main results from the present appendix are two-fold:
the existence of nontrivial solutions of the
stationarity equation \eqref{eq:app-algebraic-equation-D3-N3}
for $(D,\,N)=(3,\,3)$
and a clear diagonal/band-diagonal structure in these solutions.
The chosen values of $D$ and $N$ are obviously
far below the values \eqref{eq:D-N-F-for-IIB-matrix-model} needed for
the genuine IIB matrix model~\cite{IKKT-1997,Aoki-etal-review-1999}.
But it is an important point of principle
to have established the existence of these critical-point solutions,
even for small values of $D$ and $N$.
We can reverse the argument: assume that there were
no such critical-point  solutions for $D=3$ and $N=3$, then it would be
difficult to imagine that there could be critical-point solutions for $D=10$
and $N\gg 1$, as needed for the IIB matrix model.

%%D=N=3 APPB paper
%These particular values of $D$ and $N$ are, of course,
%far below the values \eqref{eq:D-N-F-for-IIB-matrix-model}
%needed for the IIB matrix model~\cite{IKKT-1997,Aoki-etal-review-1999}.
%Still, it is an important point of principle
%to have established the existence of solutions with full fermion dynamics,
%even for small values of $D$ and $N$.
%Let us turn the argument around: assume that there were
%no such solutions for $D=3$ and $N=3$, then it would be
%hard to believe that there could be solutions for $D=10$
%and $N\gg 1$, as needed for the IIB matrix model.
%%

A few follow-up remarks about our critical-point
calculation are as follows. First,
we were not able to obtain a nontrivial critical-point solution of
\eqref{eq:app-algebraic-equation-D3-N3}
if the sign of the Pfaffian term was reversed
(thereby removing the supersymmetry of the model).

Second, without a Pfaffian term in \eqref{eq:app-algebraic-equation-D3-N3},
there is obviously an infinity of diagonal solutions.

Third, comparing the $D=N=3$ critical solutions obtained
in the present appendix with the master-field solutions
from Ref.~\cite{Klinkhamer2021-sols-D3-N3},
we see that the randomness of the master momenta $\widehat{p}_{k}$  and
the master-noise  matrices $\widehat{\eta}^{\;\mu}_{\;kl}$
appears to have reduced somewhat the strength of the band-diagonality
in the master-field solutions [compare the $\text{O}(5)$ ratio value on the second
row of Table~\ref{tab-comparison-num-sols} with the larger nontrivial values
in \eqref{eq:app-Atilde-diagonalized-ratios}, for example].

Fourth, it is possible to interpret the bosonic critical-point solution
as  ``classical'' (cf. Refs. \cite{IKKT-1997,Steinacker2022})
and the $\{\widehat{p}_{k},\,\widehat{\eta}^{\;\mu}_{\;kl} \}$ randomness
entering the master-field equation as ``bosonic quantum fluctuations''
(the ``fermionic quantum fluctuations'' have already been
included in the effective bosonic action).
But this is only a qualitative interpretation, as there is
no small dimensionless coupling constant in the IIB matrix model
(see our previous remarks in Sec.~\ref{subsec:Conceptual-remarks}).

%\end{appendix}

%%\newpage %%tmp

\end{document}